\documentclass[twocolumn,preprint2]{aastex631}

\usepackage{placeins} 
\usepackage{amsmath}

\begin{document}

\title{COSMOS-Web: A Multi-wavelength Morphological Catalog of $\sim$780,000 Galaxies}

\author[0000-0002-8434-880X]{Lilan Yang}
\affiliation{Laboratory for Multiwavelength Astrophysics, School of Physics and Astronomy, Rochester Institute of Technology, 84 Lomb Memorial Drive, Rochester, NY 14623, USA}
\affiliation{Department of Physics, School of Physics and Electronics, Hunan Normal University, Changsha 410081, People’s Republic of China}

\author[0000-0001-9187-3605]{Jeyhan S. Kartaltepe}
\affiliation{Laboratory for Multiwavelength Astrophysics, School of Physics and Astronomy, Rochester Institute of Technology, 84 Lomb Memorial Drive, Rochester, NY 14623, USA}

\author[0000-0003-4761-2197]{Nicole E. Drakos}
\affiliation{Department of Physics and Astronomy, University of Hawaii, Hilo, 200 W Kawili St, Hilo, HI 96720, USA}

\author[0000-0002-9382-9832]{Andreas L. Faisst}
\affiliation{Caltech/IPAC, MS 314-6, 1200 E. California Blvd. Pasadena, CA 91125, USA}

\author[0009-0000-0272-5468]{Carter Flayhart}
\affiliation{Laboratory for Multiwavelength Astrophysics, School of Physics and Astronomy, Rochester Institute of Technology, 84 Lomb Memorial Drive, Rochester, NY 14623, USA}

\author[0000-0002-3560-8599]{Maximilien Franco}
\affiliation{Université Paris-Saclay, Université Paris Cité, CEA, CNRS, AIM, 91191 Gif-sur-Yvette, France}

\author[0000-0002-6610-2048]{Anton M. Koekemoer}
\affiliation{Space Telescope Science Institute, 3700 San Martin Dr., Baltimore, MD 21218, USA} 

\author[0000-0002-7303-4397]{Olivier Ilbert}
\affiliation{Aix Marseille Univ, CNRS, CNES, LAM, Marseille, France  }

\author[0000-0003-3596-8794]{Hollis B. Akins}
\altaffiliation{NSF Graduate Research Fellow}
\affiliation{The University of Texas at Austin, 2515 Speedway Blvd Stop C1400, Austin, TX 78712, USA}

\author[0000-0002-0930-6466]{Caitlin M. Casey}
\affiliation{Department of Physics, University of California, Santa Barbara, Santa Barbara, CA 93106, USA}
\affiliation{Cosmic Dawn Center (DAWN), Denmark}

\author[0000-0001-8917-2148]{Xuheng Ding}
\affiliation{School of Physics and Technology, Wuhan University, Wuhan 430072, China}

\author[0009-0003-3097-6733]{Ali Hadi}
\affiliation{Department of Physics and Astronomy, University of California, Riverside, 900 University Ave, Riverside, CA 92521, USA}

\author[0000-0003-0129-2079]{Santosh Harish}
\affiliation{Space Telescope Science Institute, 3700 San Martin Dr., Baltimore, MD 21218, USA}
\affiliation{Laboratory for Multiwavelength Astrophysics, School of Physics and Astronomy, Rochester Institute of Technology, 84 Lomb Memorial Drive, Rochester, NY 14623, USA}

\author[0000-0002-0322-6131]{Ronaldo Laishram}
\affiliation{National Astronomical Observatory of Japan, 2-21-1 Osawa, Mitaka, Tokyo 181-8588, Japan}

\author[0000-0001-9773-7479]{Daizhong Liu}
\affiliation{Purple Mountain Observatory, Chinese Academy of Sciences, 10 Yuanhua Road, Nanjing 210023, China}

\author[0000-0002-4872-2294]{Georgios E. Magdis}
\affiliation{Cosmic Dawn Center (DAWN), Denmark} 
\affiliation{DTU-Space, Technical University of Denmark, Elektrovej 327, 2800, Kgs. Lyngby, Denmark}
\affiliation{Niels Bohr Institute, University of Copenhagen, Jagtvej 128, DK-2200, Copenhagen, Denmark}

\author[0000-0002-9883-1413]{Felix Martinez III}
\affiliation{Laboratory for Multiwavelength Astrophysics, School of Physics and Astronomy, Rochester Institute of Technology, 84 Lomb Memorial Drive, Rochester, NY 14623, USA}

\author[0000-0002-9489-7765]{Henry Joy McCracken}
\affiliation{Institut d’Astrophysique de Paris, UMR 7095, CNRS, and Sorbonne Université, 98 bis boulevard Arago, F-75014 Paris, France}

\author[0000-0003-2397-0360]{Louise Paquereau} 
\affiliation{Institut d’Astrophysique de Paris, UMR 7095, CNRS, and Sorbonne Université, 98 bis boulevard Arago, F-75014 Paris, France}

\author[0000-0002-4485-8549]{Jason Rhodes}
\affiliation{Jet Propulsion Laboratory, California Institute of Technology, 4800 Oak Grove Drive, Pasadena, CA 91001, USA}

\author[0000-0002-4271-0364]{Brant E. Robertson}
\affiliation{Department of Astronomy and Astrophysics, University of California, Santa Cruz, 1156 High Street, Santa Cruz, CA 95064, USA}

\author[0000-0002-7087-0701]{Marko Shuntov}
\affiliation{Cosmic Dawn Center (DAWN), Denmark} 
\affiliation{Niels Bohr Institute, University of Copenhagen, Jagtvej 128, DK-2200, Copenhagen, Denmark}
\affiliation{University of Geneva, 24 rue du Général-Dufour, 1211 Genève 4, Switzerland}

\author[0009-0005-3133-1157]{Greta Toni}
\affiliation{University of Bologna - Department of Physics and Astronomy “Augusto Righi” (DIFA), Via Gobetti 93/2, I-40129 Bologna, Italy}
\affiliation{INAF- Osservatorio di Astrofisica e Scienza dello Spazio, Via Gobetti 93/3, I-40129, Bologna, Italy}
\affiliation{Zentrum f\"{u}r Astronomie, Universit\"{a}t Heidelberg, Philosophenweg 12, D-69120, Heidelberg, Germany}

\begin{abstract}
We present multi-wavelength morphological measurements for all galaxies in the COSMOS-Web survey, i.e., $\sim$780,000 galaxies contained in the COSMOS2025 catalog. We perform both parametric (e.g., single and double S\'ersic modeling) and non-parametric
(e.g., Gini-$M_{20}$) morphology analyses in four NIRCam bands, independently. 
Our parametric fits reveal a strong correlation between galaxy structure and star formation activity up to $z\sim4$, as evidenced by the dependence of the S\'ersic index ($n_{\text{s\'ersic}}$) and bulge-to-total ratio ($B/T$) on the position of the star formation rate–stellar mass plane.
A tight correlation between $n_{\text{s\'ersic}}$ and $B/T$ is observed.
The evolution of $n_{\text{s\'ersic}}$ and $B/T$ depends on stellar mass; for example, the median $n_{\text{s\'ersic}}$ increases from $\sim1$ at $z\sim6$ to $\sim2.5$ at $z\sim2$ for massive galaxies with $M_*>10^{10.5} M_{\odot}$, while lower mass galaxies remain $n_{\text{s\'ersic}}\sim1.2$ at all epochs.
The UV $n_{\text{s\'ersic}}$ values are systematically smaller than those in the optical, although both exhibit similar evolutionary trends.
From non-parametric analyses, we demonstrate the distribution of galaxies on the Gini–$M_{20}$ and asymmetry–concentration planes, and find that morphological classifications based on non-parametric indicators are consistent with those derived from the S\'ersic index.
The resulting catalog provides the largest and most detailed set of JWST multi-wavelength morphological measurements to date, serving as a valuable community resource for studies of structural transformation, bulge growth, and galaxy-supermassive black hole coevolution across cosmic time.

\end{abstract}


\section{Introduction}

Galaxy morphology is both a fundamental observational property and a key to understanding galaxy formation and evolution \citep[see review by][]{Conselice2014}. In the local universe, galaxies can be well classified into their Hubble types, i.e., spirals, ellipticals, and irregulars \citep{Hubble1926}. One of the key open questions is understanding when the Hubble Sequence was first established (i.e., when spirals, ellipticals, bulges, and bars first formed), and how it evolved over cosmic time. Recent James Webb Space Telescope \citep[JWST,][]{Gardner2006} observations have enabled the exploration of morphological diversity up to z$\sim$9, and many studies \citep{Ferreira2022, Ferreira2023, Kartaltepe2023, Huertas-Company2024, Huertas-Company2025} have shown that Hubble-type galaxies seem to emerge earlier than previously known from Hubble Space Telescope (HST) datasets \citep{Buitrago2013, Mortlock2013}, thanks to JWST's unprecedented sensitivity and spatial resolution. 

JWST observations reveal a larger fraction of disk galaxies at higher redshift than earlier HST observations (e.g., \citealt{Ferreira2022,Kartaltepe2023,Huertas-Company2024}), largely due to JWST's increased sensitivity. However, the exact time when the first disks formed remains uncertain. Additionally, many earlier HST studies have found that massive galaxies are typically dominated by bulges up to $z\sim3$. \cite{Huertas-Company2024} confirmed this trend using JWST data and further reported the prevalence of bulge-dominated galaxies up to $z\sim6$. Both JWST and HST observations have shown that galaxy morphologies become more irregular and disturbed at higher redshifts, with peculiar galaxies dominating at $z>4$ \citep{Huertas-Company2025}.

There is a strong correlation between star formation activity and galaxy structure, where star-forming galaxies are often disk-dominated, while quiescent galaxies are bulge-dominated \citep{Kauffmann2003, Martig2009, Wuyts2011}. A bimodality is also observed in the size-mass plane \citep{vdW2014, Yang2021, Yang2025}, where quiescent galaxies tend to be smaller than their star-forming counterparts, leading to a higher stellar mass density in the quiescent population. However, the causal relationship between morphological transformation and quenching remains debated, giving rise to two distinct scenarios: mass quenching and environmental quenching \citep{Peng2010}.
Some studies suggest that, for massive galaxies, the formation of a bulge (or the increase in central stellar surface density) precedes quenching. For example, \citet{Barro2017} found that massive galaxies likely transition to quiescence once they reach a certain density threshold. The increase in central density correlates with intense central star formation and the growth of central supermassive black holes, as the mass of galaxy bulges is tightly associated with black hole mass \citep{Kormendy2013}. On the other hand, other studies have found that massive galaxies can be quenched without undergoing significant morphological transformation \citep{vdw2011, Fudamoto2022, Cui2024}.

Since the structure of a galaxy is closely linked to its evolutionary history and star formation activity, accurate morphological measurements are essential for exploring the underlying mechanisms that shape galaxies over cosmic time.  To this end, we present a comprehensive morphological catalog for galaxies from COSMOS-Web survey, the largest area JWST imaging survey \citep[PIs: Kartaltepe \& Casey, PID: 1727][]{Casey2023}, based on the ultra-deep galaxy catalog of \cite{Shuntov2025}.
The COSMOS-Web galaxy catalog (hereafter COSMOS2025) contains over 780,000 galaxies. This work provides multiple quantitative morphological diagnostics for each of these sources, including parametric modeling based on S\'ersic profiles \citep{Sersic1968} and non-parametric image statistics.

This paper is organized as follows. In Section~\ref{sec:data}, we describe the data used for our measurements and the COSMOS2025 catalog. In Section~\ref{sec:para-measure}, we describe the methodologies for the parametric and non-parametric morphological measurements. The results from both sets of measurements are presented in Section~\ref{sec:results}. In Section~\ref{sec:comparsion}, we compare our measurements with two independent results obtained from \texttt{SourceXtractor++} \citep[\texttt{SE++};][]{Bertin2020, kummel2020} and a machine learning technique \citep{Huertas-Company2025}. Lastly, we evaluate our measurement uncertainty in Section~\ref{sec:uncertainties} and present the summary and conclusion in Section~\ref{sec:sum_and_conclusion}. We adopt the AB magnitude system \citep{Oke1983} throughout and use the \citet{Chabrier2003} initial mass function for computing stellar masses.

\section{Data}\label{sec:data}
\subsection{COSMOS-Web Imaging Survey}
The COSMOS-Web survey \citep[PIs: Kartaltepe \& Casey, PID: 1727][]{Casey2023} is the largest imaging survey selected for observations during JWST Cycle 1 and consists of contiguous 0.54 deg$^2$ Near-Infrared Camera \citep[NIRCam,][]{Rieke2023} observations in four filters (F115W, F150W, F277W and F444W) and non-contiguous 0.2 deg$^2$ Mid-Infrared Imager \citep[MIRI,][]{Wright2023} observations in F770W. The depth of the NIRCam data is measured to be 26.6--27.3 mag (F115W), 26.9--27.7 mag (F150W), 27.5--28.2 mag (F277W), and 27.5--28.2 mag (F444W) for 5$\sigma$ point sources calculated within 0.15 arcsec radius apertures. The depths of the MIRI observations are 25.33--25.98 mag calculated within 0.3 arcsec radius apertures.

The details of the NIRCam and MIRI data reduction have been presented by \cite{Franco2025} and \cite{Harish2025}, respectively. The imaging data were processed using the standard JWST Calibration Pipeline, with additional optimizations to refine the data quality. The reduction procedure, from raw uncalibrated data to the final mosaics, consists of three stages. Stage 1 performs detector-level corrections. For example, corrections for the NIRCam data include the removal of snowball events, `wisp' features, $1/f$ noise subtraction, and mitigation of some persistence effects. Following the initial corrections, Stage 2 handles image processing, resulting in calibrated science images. The final stage generates unified, science-ready mosaic images and includes several steps such as astrometric alignment, outlier detection, and resampling.

Ultimately, the COSMOS-Web survey delivers publicly accessible, science-ready mosaic images divided into 20 tiles \footnote{https://cosmos.astro.caltech.edu/page/cosmosweb}. The data are provided in the `i2d.fits' format with detailed descriptions of all included data extensions available in the JWST documentation. Additionally, each mosaic is produced at two resolutions, corresponding to pixel scales of 30 mas and 60 mas. In this work, we adopt the 30 mas mosaics to perform morphological measurements.

\subsection{The COSMOS-Web Galaxy Catalog}
COSMOS-Web's COSMOS2025 catalog provides photometry, morphology, 
photometric redshifts, and physical parameters for $\sim780,000$ galaxies \footnote{https://cosmos2025.iap.fr/catalog.html}. Here, we briefly summarize the construction process of the catalog. We refer the reader to \cite{Shuntov2025} for a detailed description. 

\subsubsection{Source Detection}
Source detection for the COSMOS-Web catalog is performed based on the detection image, which is the combination of all four NIRCam bands. The point spread function (PSF) in each filter is modeled via \texttt{PSFEx} \citep{Bertin2011} by taking all stars extracted in the field of view as input. The PSF-homogenized science mosaics in each filter are generated convolving each image to match the PSF of the F444W filter.  Noise-equalized images are then created by multiplying the PSF-homogenized science mosaics by the square root of their corresponding weight maps. The root-mean-square (RMS) of these noise-equalized images is measured by iteratively fitting a Gaussian function to the lower end of the pixel value distribution. Using the PSF-homogenized, noise-equalized images and their corresponding RMS images, signal-to-noise ratio (SNR) maps are generated for each filter. These SNR maps are then added in quadrature to produce a nominal $\chi^2$ detection image. The final detection image is taken as the square root of this $\chi^2$ image.

Using this detection image, source detection is carried using \texttt{SEP} \citep{Barbary2016}, the Python implementation of \texttt{SourceExtractor} (\texttt{SE}, \citealt{Bertin1996}). The \texttt{hot+cold} detection strategy \citep{Rix2004} is adopted to optimize detection for both bright and extended sources (cold mode), and faint, isolated sources (hot mode). In total, 784,016 sources are detected in the COSMOS2025 catalog, with 566,521 identified by cold-mode detection and 217,495 added by hot-mode detection.

In this work, we adopt the same PSFs and segmentation maps from the source detection process to perform morphological measurements. The segmentation maps are used to model companion sources and to mask out contamination during our analysis.

\subsubsection{Photometric Measurements}\label{sec:photometric-measure}
For each of the detected sources, two methods are used to measure galaxy photometry. The first method uses \texttt{SE} on the PSF-homogenized images and the other performs model-based photometry using \texttt{SE++} \citep{Bertin2020, kummel2020} on the images at their native resolution. Aperture photometry is measured for HST/ACS F814W \citep{Koekemoer2007}, the four NIRCam filters, and MIRI F770W based on circular apertures and Kron elliptical apertures. The second method is that \texttt{SE++} performs a multi-band S\'ersic model fit to measure photometry across 37 ground- and space-based filters for all detected sources.

\texttt{SE++} uses a single S\'ersic to model the galaxies, with model parameters consisting of flux, S\'ersic index, effective radius, axis ratio, and position angle. The modeling is conducted in two stages. \texttt{SE++} first fits the galaxies across the four NIRCam bands simultaneously. As such, the obtained structural parameters correspond to the averaged morphology over the 1--5 micron wavelength range. Second, \texttt{SE++} only fits flux and small coordinate offsets of galaxies in the remaining bands, while keeping the structural parameters fixed as averaged values obtained from the first step.

\subsubsection{Photometric Redshifts and Physical Properties}

Photometric redshifts and physical properties are estimated using the template-fitting code \texttt{LePHARE} \citep{arnouts_measuring_2002,ilbert_accurate_2006}.  The code utilizes a set of galaxy templates generated from the \citet{Bruzual2003} stellar population synthesis models. These templates assume six different star formation histories, include emission lines, and span a broad range of dust attenuation scenarios. In addition to galaxy templates, stellar and AGN templates are also incorporated in parallel during the fitting process. The posterior distribution of the photometric redshift is obtained, and the median value is adopted as the fiducial redshift estimate. To derive physical properties, \texttt{LePHARE} performs a second fitting run with the redshift fixed to this estimated value.

\begin{figure*}
\centering
\includegraphics[width=1.8\columnwidth]{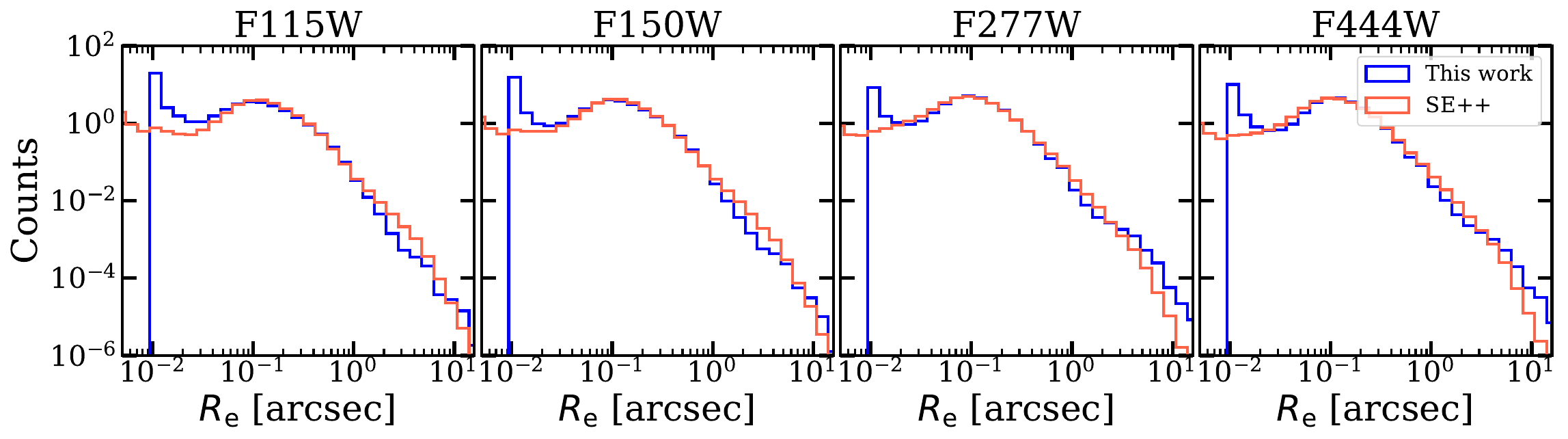}\\
\includegraphics[width=1.8\columnwidth]{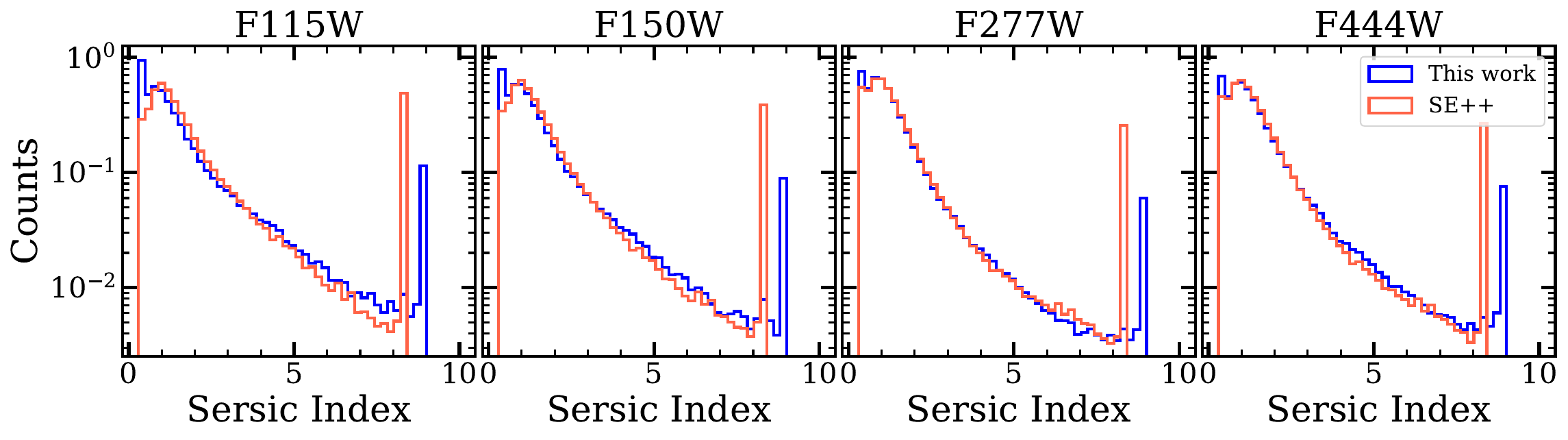}\\
\includegraphics[width=1.8\columnwidth]{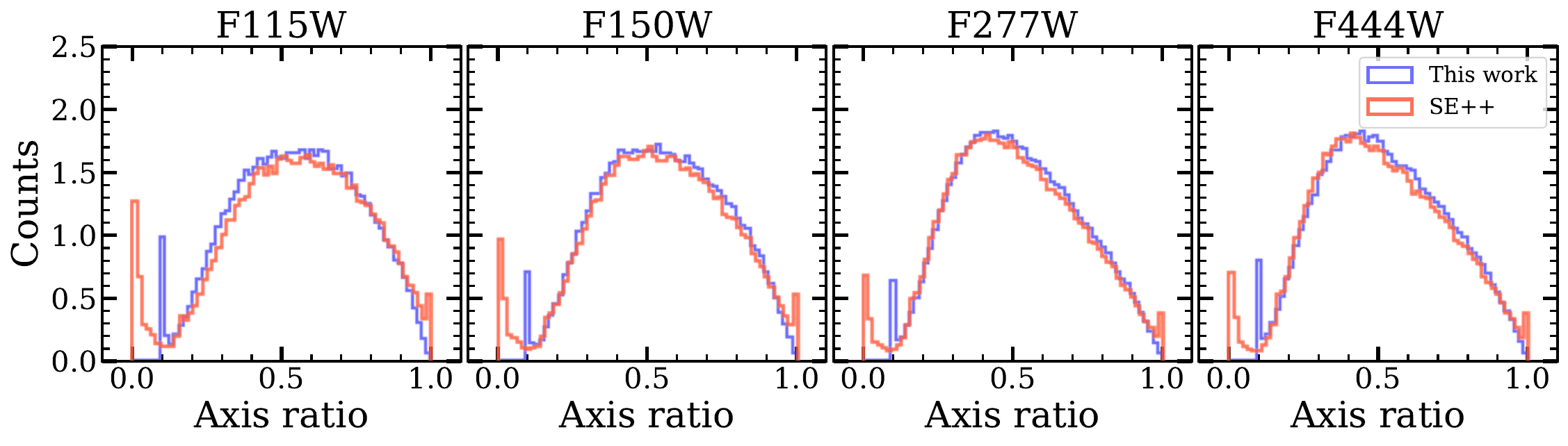}

\caption{Normalized density distribution of parameters measured from single S\'ersic modeling: half-light radius $R_{\text{e}}$, S\'ersic index, and axis ratio for sources with SNR$>20$ in the F115W, F150W, F277W, and F444W filters, respectively. The blue lines indicate measurements from \texttt{Galight} in each filter independently, while the orange lines show \texttt{SE++} results, which represent effectively averaged values over the four bands. }
\label{img:parameters-single-sersic}
\end{figure*}

\begin{figure*}
\centering
\includegraphics[width=1.8\columnwidth]{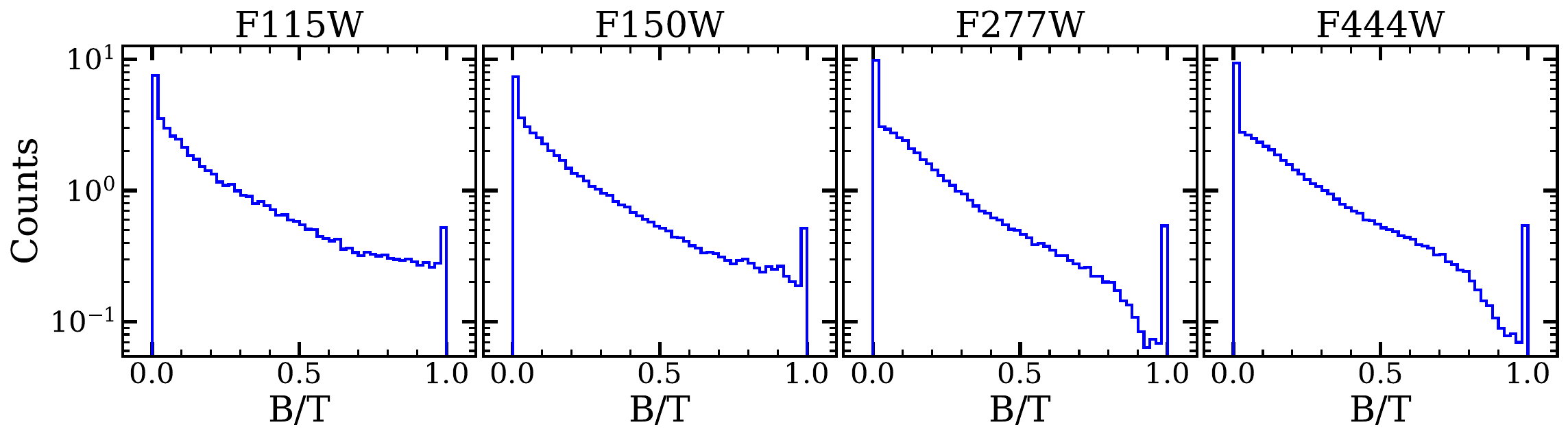}
\caption{Distribution of the bulge-to-total ($B/T$) ratio in the F115W, F150W, F277W and F444W filters for sources with SNR$>20$. Each source is fitted by a double S\'ersic model with $n_{\text{s\'ersic}}$ fixed to 1 and 4 for the disk and bulge components, respectively. }
\label{img:2sersic_b2t_distribution}
\end{figure*}

\section{Morphological Measurements}\label{sec:para-measure}

\subsection{Parametric measurements}

To construct our multi-wavelength morphological catalog, we perform two sets of measurements: parametric and non-parametric. In both cases, we use the native mosaic images with a pixel scale of 30 mas. In this section, we describe our parametric model-fitting approach using \texttt{Galight} \citep{Ding2020}, which is applied to all detected sources across all four NIRCam filters.

\subsubsection{Strategy}\label{sec:fits}
We use \texttt{Galight} to perform S\'ersic parametric fits for all sources. \texttt{Galight} is designed to perform two-dimensional model fitting of images to characterize the light distribution of galaxies, and is a wrapper package that facilitates the setup of input materials for \texttt{lenstronomy} \citep{Birrer2021}. To maximize the scientific application of the catalogs, we fit each source by three scenarios, including a single S\'ersic model, decomposing it as bulge and disk with two S\'ersic models, and modeling it by adding a point source at the center to describe the potential AGN scenarios.
Multi-band fitting diagnostic plots for all three configurations are also provided online.\footnote{For example, see single S\'ersic model results for source with ID=2026,
\url{https://cosmos2025.iap.fr/fitsmap/data/morph_plots/1sersic/2026_1sersic.pdf}.}
Our fitting strategy is as follows.

\begin{itemize}
   
\item 1. Single S\'ersic model: we use a 2D S\'ersic model to fit the light distribution of galaxies \citep{Sersic1968}:
\begin{equation}
I(r) = I_0 \exp \left[ -b_n \left(\frac{R(x,y)}{R_e}\right)^{\frac{1}{n_{\text{s\'ersic}}}} - 1\right],
\end{equation} 
where $I_0$ is the surface brightness amplitude at the half-light radius $R_{\text{e}}$, $n_{\text{s\'ersic}}$ is the S\'ersic index, and $b_n$ is a parameter dependent on $n_{\text{s\'ersic}}$. $R=\sqrt{x^2+y^2/q^2}$, where $(x,y)$ are the pixel coordinates of the source and $q$ is the axis ratio. We force the half-light radius to be between 0.01 arcsec and the radius of the cutout frame size.
The lower limit prevents fitting unrealistically small galaxies, while still allowing measurement of compact ones to be resolved, and the upper limit ensures the fitting within the field of view.
We set $n_{\text{s\'ersic}}$ between 0.3 and 9.
The light distribution of disk galaxies typically follows an exponential distribution, $n_{\text{s\'ersic}}\sim1$, while some diffuse galaxies have even lower light distribution values. To account for these factors and avoid unreasonably shallow distributions, we set the lower bound for $n_{\text{s\'ersic}}$ to 0.3.  
The light distribution of elliptical galaxies is typically $n_{\text{s\'ersic}}>2.5$, while some highly concentrated galaxies, such as the brightest galaxy clusters, can have very high light distribution values; therefore, we set the upper bound to 9.
For the axis ratio $q$, edge-on disk galaxies have $q\sim0.2-0.3$, while $q=1$ represents a perfect circular system. To ensure reasonable fits, we set $q$ to to range from 0.1 to 1.
The centroid position is allowed to vary within 2 pixels.

\item 2. Double S\'ersic model: we decompose the galaxies into two components, a disk and a central bulge, and describe each by a S\'ersic model with $n_{\text{s\'ersic}}=1$, and $n_{\text{s\'ersic}}=4$, respectively.  The constraints on the other S\'ersic parameters are the same as for the single S\'ersic model, and additional constraints have been applied. The size of the bulge must be smaller than the disk, the disk is required to have greater ellipticity, and the offset between centroids must be within 1 pixel. 

\item 3. Single S\'ersic model with a central point source: to describe the structure of potential AGN candidates, we also decompose each source into a point source with an extended S\'ersic component.  The initial centroid of the central point source is the same as the underlying  S\'ersic component and we allow it to vary within 2 pixels.

\end{itemize}

\subsubsection{Input preparation and fitting}
For each source, we obtain an image cutout from the `SCI' 2D image extension of the corresponding large tile mosaic FITS file. The size of the cutout frame (half of the length of one side) is chosen to be five times the half-light radius measured by \texttt{SE++}. To avoid poor fitting caused by a cutout frame that is too small, we require the cutout radius to be larger than 30 pixels. A noise map of the same size is also produced, read from the `ERR' extension.

A segmentation map, generated from COSMOS-Web source detection and with the same cutout size, is also provided to perform parametric and non-parametric morphology measurements. In the cutout field of view, if there is emission from another object, we use an additional S\'ersic profile to model its light and remove potential contamination from the target galaxy's extended profile. Sometimes, numerous contaminants are present within the field of view. In such cases, we sort these objects based on their distance from the central target source, modeling only the three nearest contaminating objects and masking the remaining sources.

Precise knowledge of the PSF is essential for accurate parameter estimation, and we use the same PSF produced by \texttt{PSFEx} that is used for the photometric measurements for our analysis (see \citealt{Shuntov2025} for more details).

With these settings, we execute the non-linear fitter Particle Swarm Optimization \citep[PSO;][]{pso} to minimize the model parameters. Initial guesses are taken from the results measured by \texttt{SE++}, as described in Section~\ref{sec:photometric-measure}. To further infer the measurement uncertainties of the parameters, we pass the PSO estimations to a Markov Chain Monte Carlo (MCMC) routine \citep{emcee}. We also evaluate uncertainty components not captured by MCMC via galaxy injection simulations, as described in Section~\ref{sec:uncertainties}.

\subsection{Non-parametric measurements}\label{sec:non-para-measure}

Parametric fitting is a powerful method for modeling galaxy structure, but it assumes that a galaxy's light profile follows a specific, smooth, and symmetric model. This assumption can break down for irregular, merging, or disturbed galaxies. To complement our parametric fits, we also perform non-parametric morphological measurements, offering a model-independent approach. Without assuming any specific light distribution, these methods can effectively characterize galaxies with complex or irregular structures \citep{Abraham1994, Schade1995, Conselice2003, Lotz2004, Barbary2016}.

We adopt the \texttt{Python} package \texttt{Statmorph} \citep{RodriguezGomez2019} to perform the non-parametric measurements, including the Gini--$M_{20}$ \citep{Lotz2004, Lotz_2008} and concentration--asymmetry--smoothness (CAS) statistics \citep{Conselice2003}. The Gini coefficient quantifies inequality in a light distribution, with a value of 1 indicating that all flux is concentrated in a single pixel, and 0 corresponding to a perfectly uniform brightness distribution. The $M_{20}$ statistic measures the second-order moment of a galaxy's brightest regions (containing 20\% of the total flux) relative to the total second-order central moment. It reflects how the galaxy's light is spatially distributed relative to the galaxy center. The CAS statistics describe the degree to which a galaxy’s light is centrally concentrated, how symmetric the galaxy is with respect to a 180 degree rotation, and the fraction of light contained in small-scale structures, respectively.

\texttt{Statmorph} requires input including cutout images, a segmentation map, a weight map or gain, a mask, and a PSF. We use the same data as for the parametric measurements. As output, the package returns non-parametric measurements along with quality flags indicating good or bad measurements. Quality flags can take the values 0 (good), 1 (suspect), 2 (bad), or 4 (catastrophic). In the following analysis, we only use results flagged as 0.

\section{Results}\label{sec:results}
This section presents the results of the morphological measurements derived from both parametric and non-parametric methods. We discuss the evolution of various morphological parameters with redshift and their dependence on stellar mass and wavelength. Throughout this and the following sections, we focus on sources with SNR $>20$ in each filter and \texttt{warn\_flag=0,2,3}, and \texttt{flag\_star\_hsc=0}. The SNR is calculated as the ratio between flux density and its error using the model fluxes in the COSMOS2025 catalog. The latter two parameters are provided in the COSMOS2025 catalog and these selections ensure that the selected sources are generally secure and their photometry is not affect by nearby bright stars in the ground-based imaging \citep[see more details in][]{Shuntov2025}.

\subsection{Distribution of parameters obtained from Single S\'ersic modeling}
The normalized density distributions of measured structural parameters from the single S\'ersic model in F115W, F150W, F277W, and F444W from \texttt{Galight} are shown in Figure~\ref{img:parameters-single-sersic}, along with a comparison to results obtained from \texttt{SE++} for the same sources. Across the four filters, the half-light radius ($R_{\rm e}$) distributions show a bump around 0.1 arcsec, with a noticeable peak at 0.01 arcsec, corresponding to unresolved, point-like sources that reach the lower boundary set in \texttt{Galight} measurements. The \texttt{SE++} results show a broadly consistent trend, although the distribution at the small-size end is flatter. This difference arises because the \texttt{SE++} lower limit on half-light radius is set as a fraction of the Kron radius \citep{Shuntov2025}, whereas \texttt{Galight} imposes a hard cutoff.

The S\'ersic index distributions are strongly right-skewed, with a peak at $n_{\text{s\'ersic}}=1$ and a gradual decline toward higher values with additional clustering near both the lower and upper boundaries, indicating that low S\'ersic index  profiles dominate the sample while high S\'ersic index profiles are progressively rarer. The distributions are similar across the four filters. In addition, the distributions of axis ratios are approximately normal, although the peak skews toward lower values, $\sim0.4$, particularly in the F277W and F444W filters. In the short-wavelength filters (F115W and F150W), the peaks are less pronounced compared to the long-wavelength filters (F277W and F444W), suggesting a wavelength-dependent structural variation. Similarly, the distributions of S\'ersic index and axis ratio obtained from \texttt{Galight} and \texttt{SE++} are generally in good agreement. A more detailed comparison between them is presented in Section~\ref{sec:gl_vs_se}.

\subsection{Distribution of Bulge-to-Total Ratio}
One of the most important structural parameters derived from double S\'ersic modeling is the bulge-to-total ($B/T$) light ratio. We present the distributions of $B/T$ across each of the four filters in Figure~\ref{img:2sersic_b2t_distribution}. The histograms exhibit a steeply declining distribution with a peak at low $B/T$ values, indicating that the majority of galaxies are disk-dominated with $B/T<0.25$. The overall trends are qualitatively consistent among the four filters. Interestingly, the distributions resemble those of the S\'ersic index obtained from single S\'ersic modeling as shown in Figure~\ref{img:parameters-single-sersic}, and their correlation is further examined in Section~\ref{sec:ns-b2t}.

\begin{figure*}
\centering
\includegraphics[width=1.8\columnwidth]{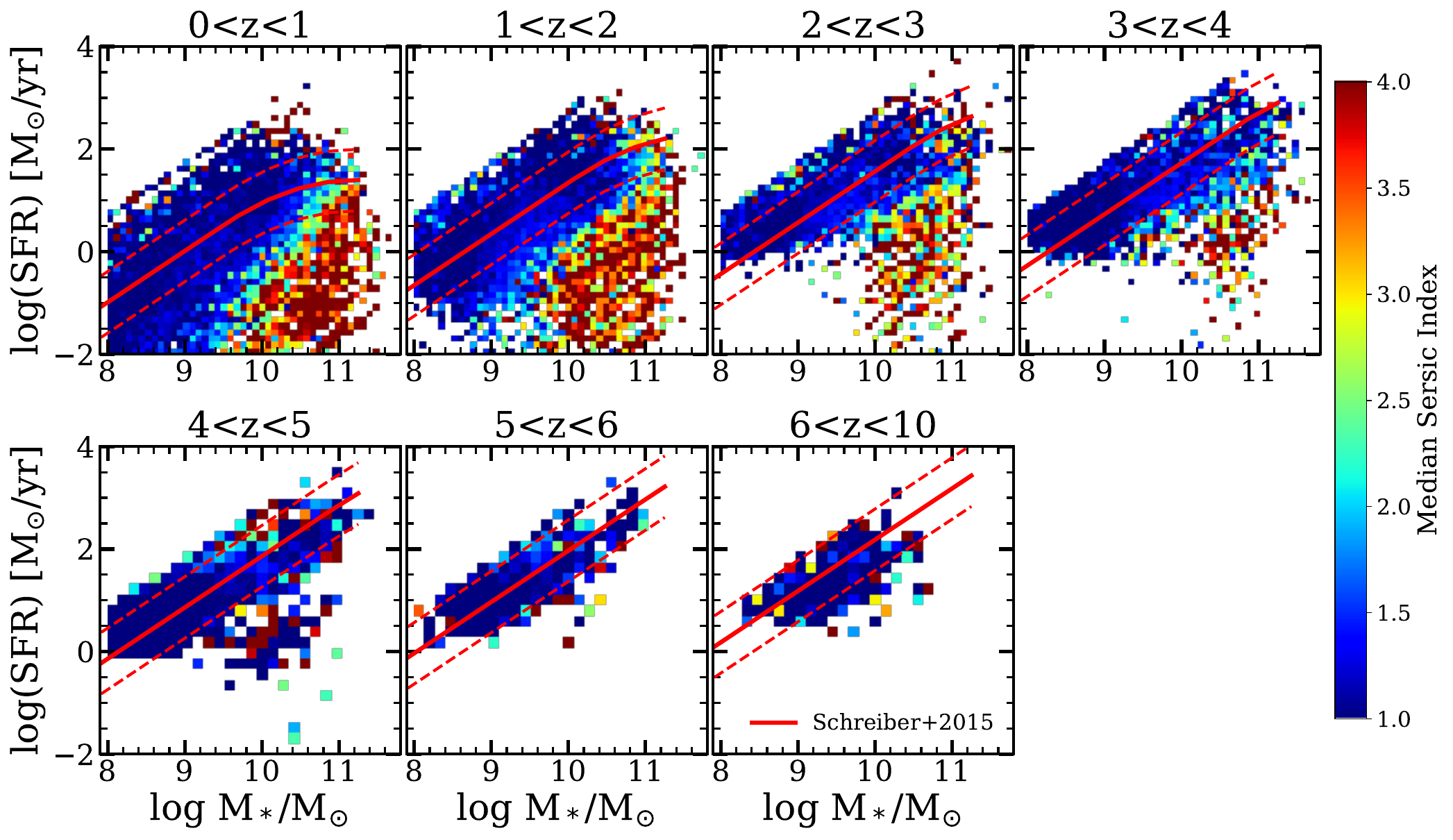}

\caption{Distribution of the S\'ersic index in the SFR-M$_{\star}$ plane at $0<z<10$ for sources with SNR$>20$. The data points are color-coded by the median value of their rest-frame optical $n_{\text{s\'ersic}}$ in each bin, with redder colors corresponding to higher values. The red lines indicate the star-forming MS adopted from \citet{Schreiber2015}, and dashed lines represent the values 4 times above and below the MS \citep{Rodighiero2011}.}
\label{img:main-sq-serisc}
\end{figure*}

\begin{figure*}
\centering
\includegraphics[width=1.8\columnwidth]{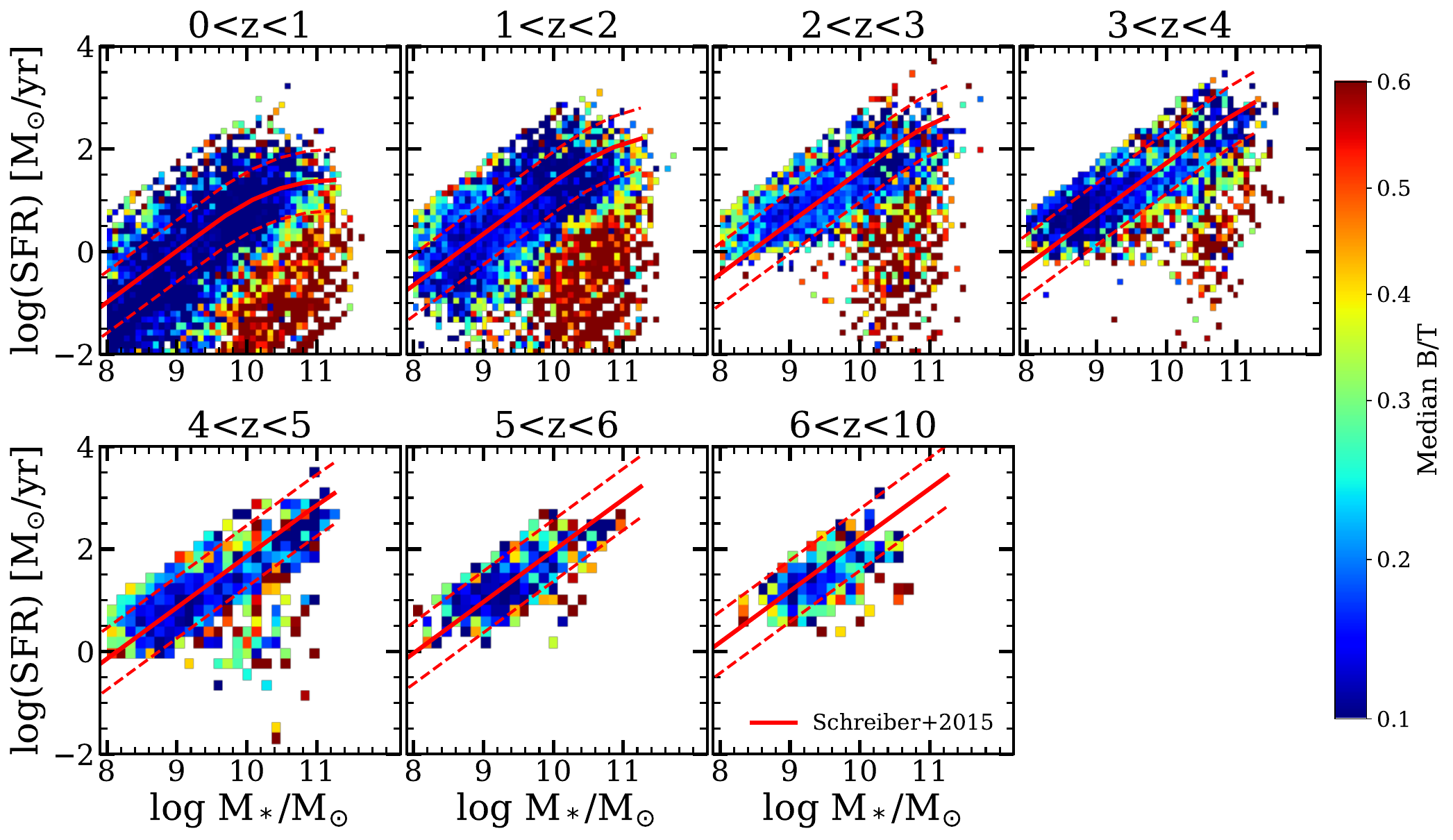}

\caption{Distribution of the bulge-to-total ratio ($B/T$) in the SFR-M$_{\star}$ plane at $0<z<10$ for sources with SNR$>20$. The data points are color-coded by the median value of $B/T$ in each bin, with redder colors corresponding to higher values. The red lines are the same as presented in Figure~\ref{img:main-sq-serisc}.}
\label{img:main-sq-b2t}
\end{figure*}

\subsection{Distribution of $n_{\text{s\'ersic}}$ on SFR--M$_{\star}$ plane}

The COSMOS2025 catalog provides photometric redshifts and physical parameters, such as stellar mass (M$_{\star}$) and star formation rate (SFR), which allow us to investigate the correlation between galaxy structure and stellar populations. We adopt the median value of the best-fit obtained by \texttt{LePHARE} for these physical parameters to conduct the following analysis. We note that there are galaxies at $z>6$ with highly uncertain high estimated stellar masses ($>10^{11}-10^{12}M_{\odot}$) contained in the COSMOS2025 catalog, which may cause some potential issues, as they might exceed values allowed within the $\Lambda$CDM framework. \citet{Lovell2023} presented the maximum stellar mass expected in $\Lambda$CDM cosmology as a function of redshift based on extreme value statistics for sources in the COSMOS-Web field, see also discussion in \citet{Franco2025b}. We thus exclude the sources with estimated mass beyond $1\sigma$ of the predicted maximum stellar mass.

In Figure~\ref{img:main-sq-serisc}, we show the dependence of galaxy structure $n_{\text{s\'ersic}}$ on position in the SFR-M$_{\star}$ plane and its evolution with redshift up to $z=10$. 
The presented $n_{\text{s\'ersic}}$ values are measured in the corresponding rest-frame optical band for each redshift bin.
The red lines indicate the star-forming main sequence (MS) adopted from \citet{Schreiber2015}, where the Salpeter IMF was used, and here we have corrected it to Charier IMF following \citet{Bernardi2017}. 
Galaxies that lie above the main sequence, are typically referred to as starburst galaxies, while galaxies that lie below are quiescent galaxies or in transition. 
The MS parameterization is constrained by star-forming galaxies data at $z<3.5$, but its extrapolation to higher redshifts shows an agreement with recent measurements, such as \citet{Khusanova2021, Clarke2024, Cole2025}.

The distribution of galaxies with exponential profiles ($n_{\text{s\'ersic}}\sim1$) follows a similar distribution in the SFR-M$_{\star}$ plane as the star-forming MS, so called ``structural MS'' \citep{Wuyts2011}. Furthermore, the diagrams generally show a smooth and continuous variation of $n_{\text{s\'ersic}}$ along the direction perpendicular to the MS at least up to $z\sim4$, where the MS star-forming galaxies have $n_{\text{s\'ersic}}\leq1$, and the quiescent galaxies below the MS have  $n_{\text{s\'ersic}}>2.5$. Additionally, the galaxies located beyond four times above the MS and those occupying the top-right tip of the MS exhibit a range of $n_{\text{s\'ersic}}$ values (e.g., median $n_{\text{s\'ersic}}$ varying from 1 to 4), with some comparable to those of quiescent galaxies. Those more compact star-forming galaxies could be the remnants of galaxy mergers \citep{Toomre1972}. A gas-rich galaxy merger can trigger intense star formation in the nucleus, leading to the depletion of gas and formation of a central stellar bulge, resulting in a higher S\'ersic index. With increased central stellar density, those galaxies are likely to transition to quiescent galaxies once they reach a certain stellar density threshold  \citep[e.g., see][]{Barro2017, Yang2025}.

However, the correlation between $n_{\text{s\'ersic}}$ and position in the SFR-M$_{\star}$ plane becomes ambiguous at $4<z<5$, where galaxies with $n_{\text{s\'ersic}}\leq1$ are visible below the MS, though the number density of this quiescent population is low at this redshift range \citep{Shuntov2025morpho}. We estimate that more than half of galaxies with $\log(M_{*}/M_{\odot})>9.5$ below the MS have $n_{\text{s\'ersic}}\leq1$ whereas the fractions are only 20\%-35\% at lower redshift bins, suggesting that quiescent populations are disk-dominated \citep{vdw2011} or irregulars at high redshifts. 
We note that we do not apply rigorous criteria to select quiescent galaxies, but other recent work that adopt strict criteria, such as color-color diagrams and specific star formation rate thresholds, show similar results that high-redshift quiescent galaxies at $z<3$ have $n_{\text{s\'ersic}}<2$ \citep[e.g.,][]{Ito2024}, and the fraction of bulge-dominated quiescent galaxies defined as $n_{\text{s\'ersic}}>2.5$ is low \citep[e.g.,][]{Yang2025}.

\subsection{Distribution of $B/T$ on SFR-M$_{\star}$ plane} 
In Figure~\ref{img:main-sq-b2t},
we further investigate the dependence of $B/T$ on  position in the SFR-M$_{\star}$ plane, and $B/T$ values are measured in the corresponding rest-frame optical band for each redshift bin.
Across all redshifts, galaxies located on the MS have a low $B/T$ ratio (i.e,  $B/T<0.2$), while those blow the MS have higher $B/T$ values. The trend is most pronounced at $z < 4$, where a clear structural separation is visible between star-forming and quiescent populations. For galaxies located below the star-forming MS (i.e., quiescent systems), we observe a clear bimodal distribution in their structural properties, especially at $z<2$. Low-mass galaxies ($\log(M_{*}/M_{\odot})<9.5$) are predominantly characterized by $B/T<0.2$ values while more massive systems ($\log(M_{*}/M_{\odot})>10.5$) are dominated by $B/T>0.5$. This bimodality suggests the presence of two distinct quenching pathways: environmental quenching, which primarily affects low-mass and disk-dominated galaxies, and mass quenching, which dominates in massive and bulge-dominated systems \citep{Peng2010}. Consistent with this picture, \citet{Shuntov2025morpho} report a similar trend in the stellar mass function (SMF) of quiescent galaxies across different $B/T$ bins. They also find that there is an upturn in the quiescent SMF at the low-mass end, providing evidence for the onset of environmental quenching as early as 
$z\sim3$.

The above results show that a strong correlation between star formation activity and galaxy structure was already in place since $z\sim4$, confirming the presence of such a correlation previously reported in the literature at lower redshift $z<3$, such as \citep{Kauffmann2003, Martig2009, Wuyts2011,Barro2017, Dimauro2022}. 
However, the clear correspondence between galaxy structure and stellar population is ambiguous at $z>4$.
Interestingly, the emergence of the Hubble sequence is also at $z\sim4$ as reported by \citet{Huertas-Company2025} where they found the abundance of Hubble types galaxies (spheroids, disk-dominated, and bulge-dominated systems) is low at $z>4$, and the galaxy population is dominated by non-Hubble types (peculiars and compacts).


\begin{figure*}
\centering
\includegraphics[width=1.8\columnwidth]{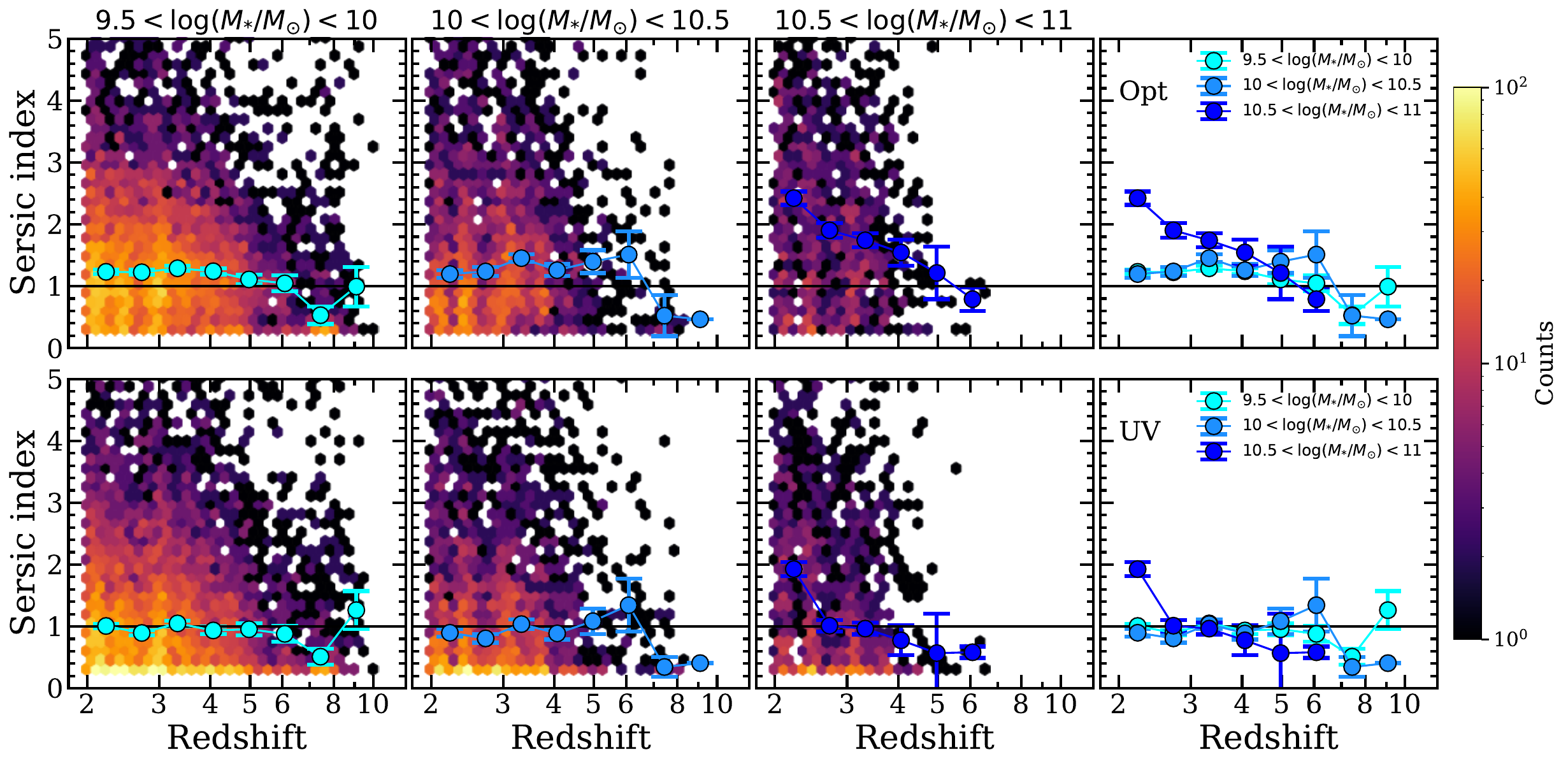}

\caption{Evolution of S\'ersic index at $2<z<10$ in the rest-frame optical (top) and UV (bottom), divided into three stellar mass bins $9.5<\log(M_{*}/M_{\odot})<10$, $10<\log(M_{*}/M_{\odot})<10.5$, and $10.5<\log(M_{*}/M_{\odot})<11$. To make a fair comparison between the two wavelength regimes and ensure a sufficient number of data points, we require sources to have SNR$>5$ in both the optical and UV. The filled symbols represent the median value, while the error bars denote the statistical uncertainties computed as $1.253\sigma/\sqrt{N}$, where $N$ is the number of galaxies and $\sigma$ is the standard deviation of the distribution within each bin. The black reference lines indicate $n_{\text{s\'ersic}}=1$.}
\label{img:ns_z_mass_uv_op}
\end{figure*}

\begin{figure*}
\centering
\includegraphics[width=1.8\columnwidth]{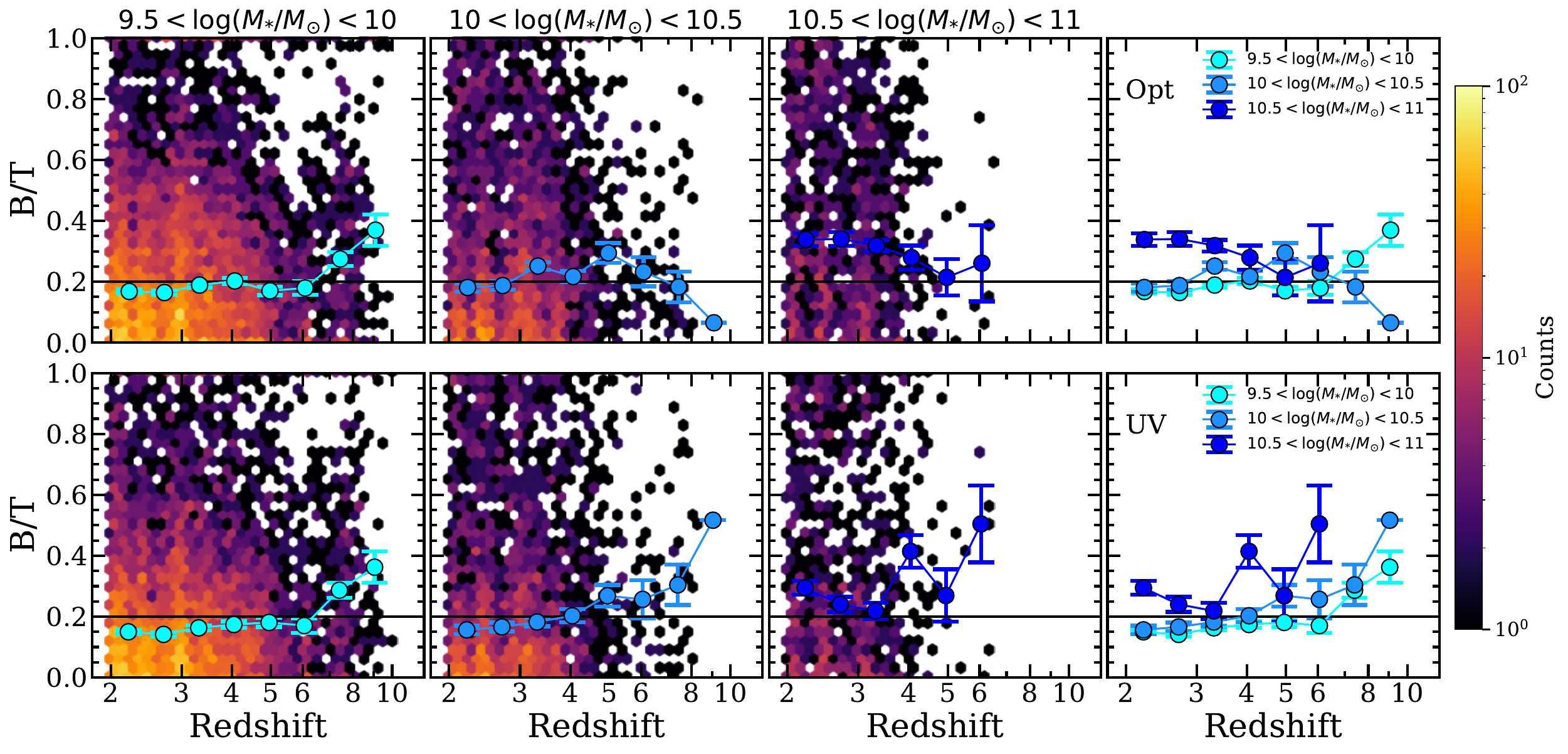}
\caption{Evolution of bulge-to-total ($B/T$) ratio at $2<z<10$ in the rest-frame optical (top) and UV (bottom), divided into three stellar mass bins $9.5<\log(M_{*}/M_{\odot})<10$, $10<\log(M_{*}/M_{\odot})<10.5$, and $10.5<\log(M_{*}/M_{\odot})<11$. For a fair comparison between the two wavelengths and to ensure a sufficient number of data points, we require sources to have SNR$>5$ in both filters. The colors and symbols are the same as shown in Figure~\ref{img:ns_z_mass_uv_op}. }
\label{img:b2t_z_mass_uv_op}
\end{figure*}

\subsection{Evolution of the S\'ersic Index}
We now investigate the evolution of the S\'ersic index $n_{\text{s\'ersic}}$ with redshift. In addition, we also examine its wavelength dependence, i.e., rest-frame optical versus rest-frame UV. 
Since the COSMOS-Web NIRCam images were observed in four filters, our analysis is restricted to the redshift range $2<z<10$. Following \citet{Yang2025}, we select F150W (F115W), F277W (F150W), and F444W (F277W) bands to represent the rest-frame optical (UV) structure for galaxies at $2<z<3$, $3<z<6$, and $6<z<10$, respectively.  
Figure~\ref{img:ns_z_mass_uv_op} shows the evolution of $n_{\text{s\'ersic}}$ in three stellar mass bins, $9.5<\log(M_{*}/M_{\odot})<10$, $10<\log(M_{*}/M_{\odot})<10.5$, and $10.5<\log(M_{*}/M_{\odot})<11$. First, we observe a clear dependence of $n_{\text{s\'ersic}}$ evolution on stellar mass. In the rest-frame optical (top panels), the median $n_{\text{s\'ersic}}$ increases significantly from $n_{\text{s\'ersic}}\leq1$ at $z\sim6$ to $n_{\text{s\'ersic}}\sim2.5$ at $z\sim2$, for the most massive galaxies $10.5<\log(M_{*}/M_{\odot})<11$. In contrast, the two lower stellar mass bins show little evolution, with the median values remaining $n_{\text{s\'ersic}}\sim1.3$. This stellar mass-dependent evolution has also reported by \citet{Martorano2025} at lower redshifts ($z < 2.5$), where they found that the median $n_{\text{s\'ersic}}$ increases significantly for the most massive galaxies with $\log(M_{*}/M_{\odot}) > 11$, and the evolution is subtle or even absent for less massive galaxies.

In the rest-frame UV (bottom panels), the median values of $n_{\text{s\'ersic}}$ are slightly smaller than in the optical at $z<6$. A similar dependence of $n_{\text{s\'ersic}}$ on wavelength has been reported by \citet{Vulcani2014, Quilley2025}. The evolution shows a similar trend, suggesting a lack of strong wavelength dependence. We also notice that at higher redshift ($z>4$), the median value of $n_{\text{s\'ersic}}$ tends to decrease in both the UV and optical, indicating the prevalence of irregular galaxies, as also seen by in \citet{Huertas-Company2025}. They found that irregular galaxies, which dominate at lower stellar masses, are more common at higher redshifts. More than half of the galaxies are irregulars at $z>3$ for galaxies with $\log(M_{*}/M_{\odot})<10.5$. They also report that irregular galaxies have a very similar distribution of S\'ersic index but with flatter profiles, leading to a lower median $n_{\text{s\'ersic}}$ as shown in this work.

\subsection{Evolution of the Bulge-to-Total Ratio}
We further examine the evolution of $B/T$ ratio and its dependence on stellar mass. Figure~\ref{img:b2t_z_mass_uv_op} illustrates its evolution and is separated into three stellar mass bins, in both rest-frame optical (top panels) and UV (bottom panels). 
At $2<z<5$, we observe a dependence of $B/T$ evolution on stellar mass similar to that of $n_{\text{s\'ersic}}$.
For example, in the rest-frame optical, the $B/T$ ratio remains almost unchanged at $2<z<5$ for galaxies at two lower mass bins. But for the massive galaxies $10.5<\log(M_{*}/M_{\odot})<11$, it increases approximately from 20\% at $z\sim5$ to 35\% at $z\sim2$. At $z>5$, surprisingly, we find a upturn of $B/T$ ratio for galaxies with $9.5<\log(M_{*}/M_{\odot})<10$ in the rest-frame optical as well as all three mass bins in the rest-frame UV, which is likely due to fitting artifacts rather than genuine structural properties. At $z>5$, most galaxies are irregulars with disturbed morphologies, and well-defined bulge-disk systems are largely absent \citep{Huertas-Company2025}. Nevertheless, when a double S\'ersic decomposition is applied, the fitting algorithm artificially assigns a central component to reproduce the concentrated light in the galaxy core, even in the absence of a physically distinct bulge. As a result, we may overestimate the bulge contribution, leading to inflated $B/T$ values. 

Therefore, we conclude that the $B/T$ ratio increases from $z=5$ to $z=2$ for the most massive galaxies in the rest-frame optical and UV, and its evolution strongly depends on stellar mass. 
We also note that the evolution in the rest-frame UV is less pronounced at $z>3$.

\begin{figure*}
\centering
\includegraphics[width=1.8\columnwidth]{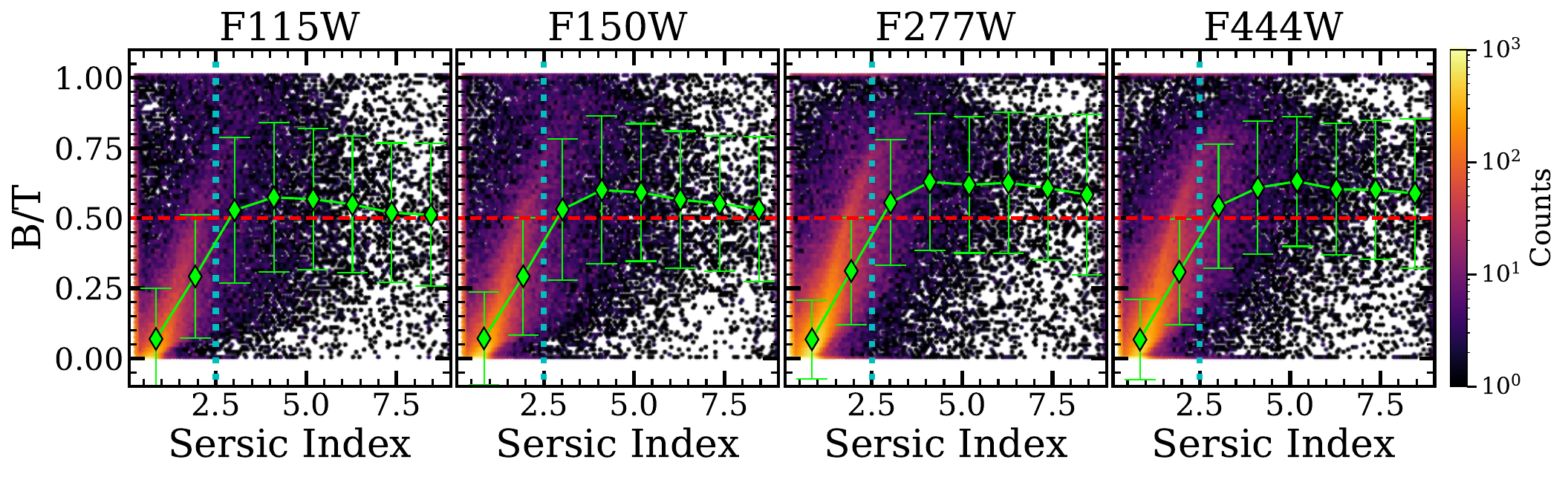}
\caption{Distribution of the bulge-to-total ($B/T$) ratio for sources as a function of the $n_{\text{s\'ersic}}$ obtained from single S\'ersic fits for sources with SNR$>20$. The lime diamonds indicate the median $B/T$ values in each bin, with error bars representing the standard deviation. The red and cyan lines mark the threshold of $B/T=0.5$ and $n_{\text{s\'ersic}}=2.5$, which are commonly used to separate bulge- and disk-dominated systems.}
\label{img:parameters-2-sersic}
\end{figure*}

\subsection{Correlation Between S\'ersic Index and $B/T$}\label{sec:ns-b2t}

Our results have demonstrated the similarity between the S\'ersic index and $B/T$, in terms of their general distribution, their distribution on  the SFR--M$_{\star}$ plane, and their evolution with redshift. Due to this similarity, the S\'ersic index is widely used as a proxy for $B/T$ \citep{Fisher2008, Weinzirl2009, Gargiulo2022}. We further check the $B/T$ ratio as a function of the S\'ersic index derived from single S\'ersic modeling across each of the four filters in Figure~\ref{img:parameters-2-sersic}. A clear positive correlation is observed where galaxies with higher S\'ersic indices tend to have higher $B/T$ ratios, although the scatter is substantial. The median value of $B/T$ is approximately 10\% for galaxies with $n_{\text{s\'ersic}}<1$, where disk galaxies dominate the population, and increases to 50\% as the $n_{\text{s\'ersic}}$ reaches 2.5, where bulges are dominant. However, the correlation becomes flat for $n_{\text{s\'ersic}}>2.5$, with the  median $B/T$ ratio staying constant with increasing $n_{\text{s\'ersic}}$. A similar correlation has also been reported from recent simulations (e.g., \citealt{Tang2025}). Moreover, this correlation is consistently seen across all four bands, supporting the robustness of the $n_{\text{s\'ersic}}$ as a structural indicator.

\subsection{Distribution of S\'ersic Index with Point source Decomposition}

\begin{figure*}
\centering
\includegraphics[width=1.8\columnwidth]{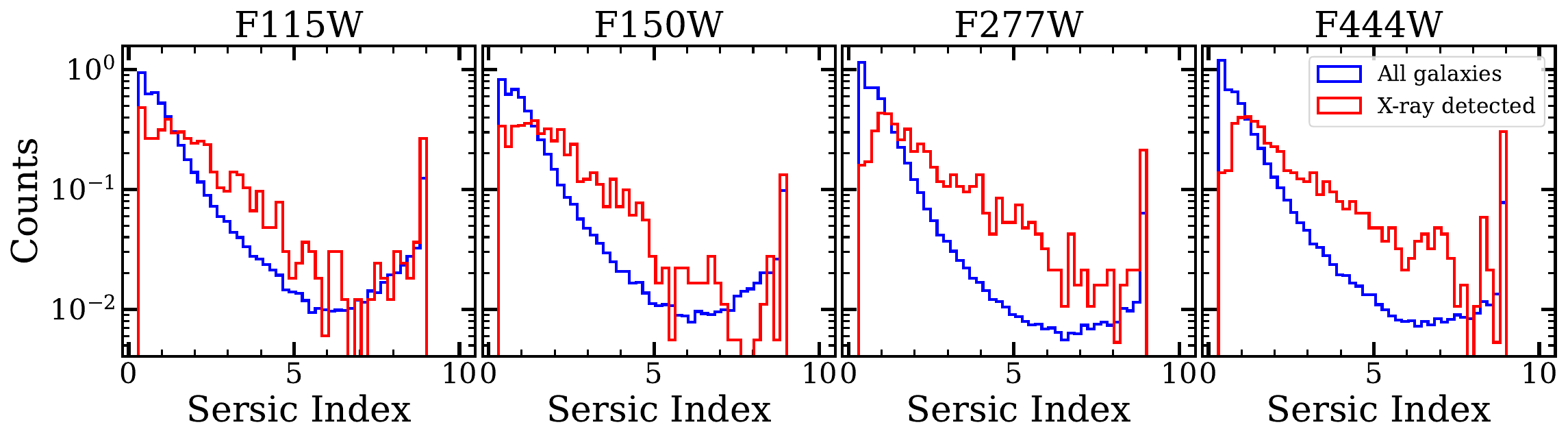}
\caption{Distribution of the S\'ersic index obtained from the decomposition of a Single S\'ersic and a central point source in the F115W, F150W, F277W and F444W filters for sources with SNR$>20$. The blue and red lines demonstrate the distribution of  
all and X-ray detected sources, respectively.
}

\label{img:parameters-ps-1sersic}
\end{figure*}

Lastly, we present the distribution of S\'ersic index obtained from the point-source decomposition as shown in Figure~\ref{img:parameters-ps-1sersic}. 
The blue lines represent the results of all sources with SNR$>20$, showing that the S\'ersic index distributions are right-skewed, similar as that  obtained from Single S\'ersic modeling. 
One of the motivations for adopting this decomposition strategy is to enable the study of host galaxies of point-like sources, such as AGNs.
Here, we demonstrate the S\'ersic index distribution of AGNs hosts as an example.
There are 1,159 X-ray detected sources with \texttt{flag\_chandra=1} in the COSMOS2025 catalog and we consider them as potential candidates for AGN. 
We show the 
distribution of the S\'ersic index obtained from decomposition for X-ray AGN hosts (red lines) in Figure~\ref{img:parameters-ps-1sersic}. 
The distribution of X-ray AGN hosts is distinct from that of all galaxies.
It has a broader distribution with a pronounced preference for a higher S\'ersic index ($n_{\text{s\'ersic}}>2$), rather than the significantly skewed distribution of the general galaxy population. We note that host measurements are performed by simultaneously modeling the central AGN as a point source, so the higher S\'ersic index is not attributed to the central bright source.

The prevalence of AGN in high S\'ersic index systems supports the scenario where bulge growth correlates with AGN activity. Such a connection aligns with models of co-evolution between the supermassive black holes (SMBHs) and their host galaxies, where dynamical processes (e.g., violent disk instabilities or mergers) funnel gas toward the nucleus, triggering star formation and a rapid phase of SMBH growth efficiently \citep{Hopkins2006, Hopkins2008a, Hopkins2008b, Georgakakis2009}. As the SMBH grows and enters an active phase, it begins to exert feedback on its host galaxy, in terms of quenching star formation and morphological transformation. The prevalence persists across all four filters, though the difference between AGN hosts and the general population is slightly more pronounced at longer wavelength, F277W and F444W.

\begin{figure*}[h]
\centering
\includegraphics[width=1.8\columnwidth]{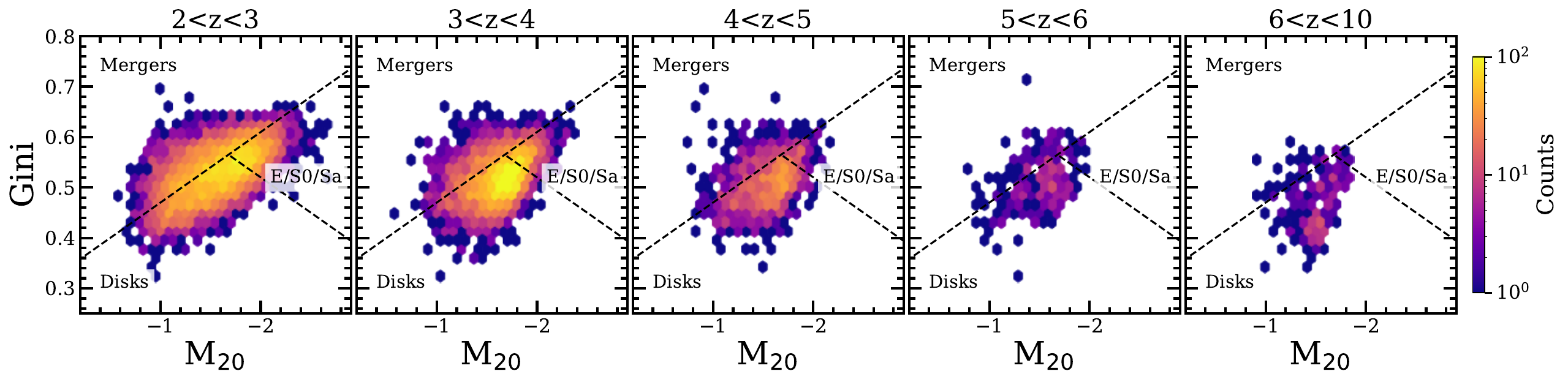}
\includegraphics[width=1.8\columnwidth]{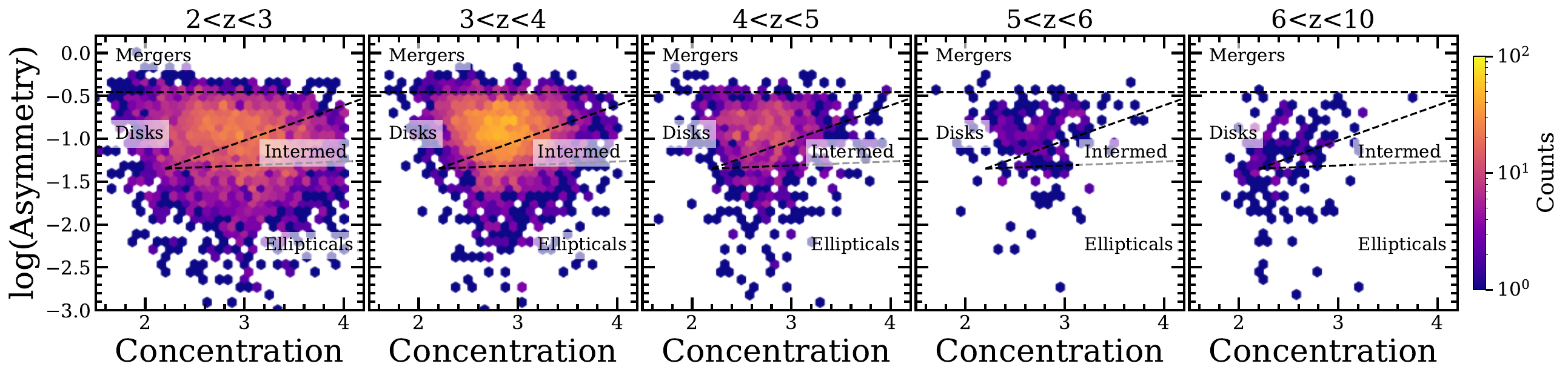}
\caption{The rest-frame optical Gini-M$_{20}$ (top) and asymmetry-concentration (bottom) planes for all galaxies with stellar mass $\log(M_*/M_\odot)>9.5$ and SNR$>20$ across redshift range $2<z<10$.
In the Gini–M$_{20}$ plane, the dashed lines indicate classical empirical divisions from \citet{Lotz2004, Lotz_2008}, which distinguish between mergers, disks (Sb--dI) and elliptical galaxies (E/S0/Sa). 
In the asymmetry–concentration plane, the dashed lines present the classification boundaries that separate disk galaxies, elliptical galaxies, intermediate types, and major mergers \citep{Bershady2000, Conselice2003}. 
}
\label{img:parameters-cas}
\end{figure*}

\begin{figure*}
\centering
\includegraphics[width=1.8\columnwidth]{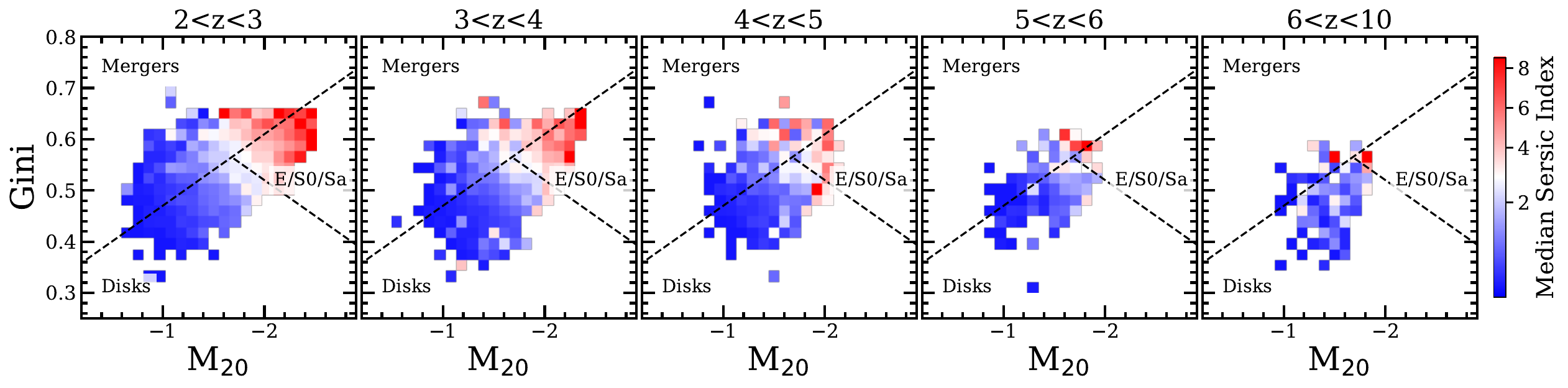}
\includegraphics[width=1.8\columnwidth]{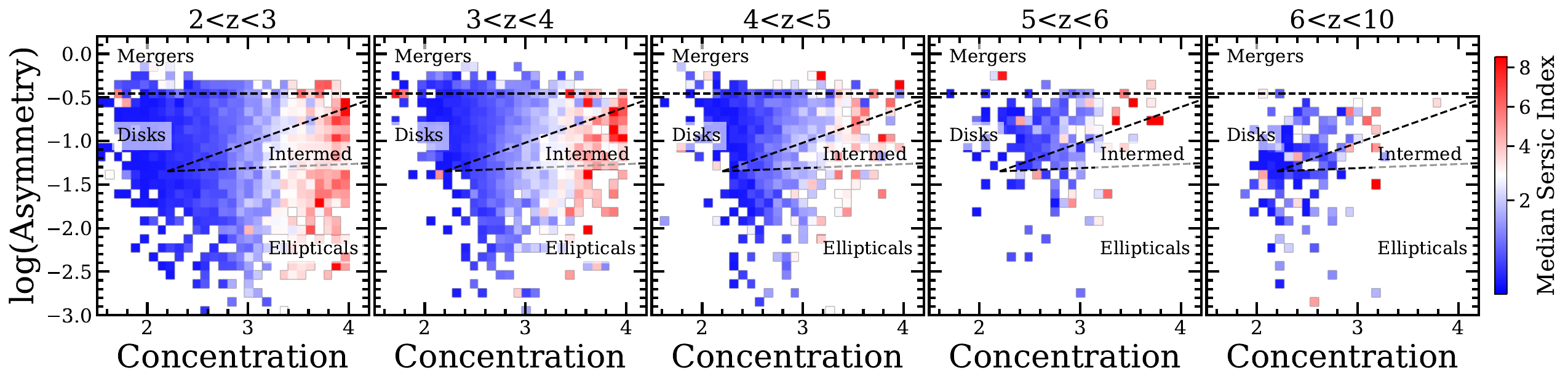}
\caption{Same as Figure~\ref{img:parameters-cas} with the color-coded by the median S\'ersic index value in each bin.}
\label{img:parameters-cas-nsersic}
\end{figure*}

\begin{figure*}
\centering
\includegraphics[width=1.8\columnwidth]{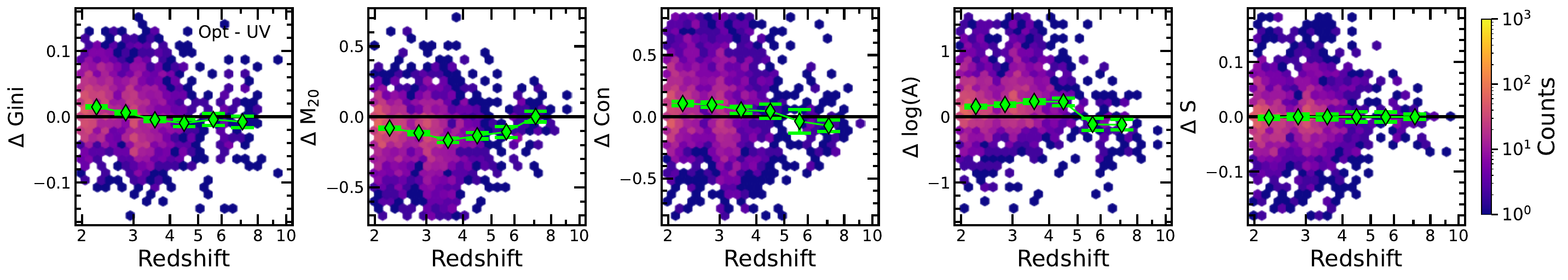}
\caption{The variation of Gini, M$_{20}$, concentration, asymmetry and smoothness between the rest-frame optical and UV as a function of redshift for sources with SNR$>5$ in both corresponding bands. 
The filled circles represent the median value, with the error bars denoting their statistical uncertainties. 
}

\label{img:delta_gmc_z}
\end{figure*}

\subsection{Non-parametric fitting results}
In addition to parametric fittings, we also employed \texttt{Statmorph} to measure non-parametric morphological statistics across all NIRCam filters. 
Here, we present results only for sources with reliable \texttt{Statmorph} measurements, i.e., those with a quality flag equal to 0 (good).
For example, there are approximately 25\% and 90\% measurements with quality flag values $=0$ and $<=1$ in the F277W band, respectively.
Figure~\ref{img:parameters-cas} shows the rest-frame optical distribution of galaxies with stellar mass $\log(M_*/M_\odot)>9.5$ in both the Gini–M$_{20}$ and asymmetry–concentration planes, two widely used diagnostics for classifying galaxy morphologies and identifying mergers.
In the Gini–M$_{20}$ planes (top), we show the classical empirical divisions from \citet{Lotz2004, Lotz_2008}, which distinguish between mergers, disk galaxies, and ellipticals. 
While in the asymmetry–concentration planes (bottom), we include classification boundaries that separate disk galaxies, elliptical galaxies, intermediate types, and mergers \citep{Bershady2000, Conselice2003}.
The classification based on the two empirical methods is somewhat distinctive, for example, there are $\sim$30\% galaxies identified as mergers in the Gini–M$_{20}$ plane while only a few galaxies are classified as mergers in the asymmetry–concentration plane.
The mergers identified in the Gini–M$_{20}$ plane distribute sparsely in the `Disks' portion in the asymmetry–concentration plane.

We first examine the correlation between non-parametric statistics and S\'ersic parameters measured via single S\'ersic modeling.
To illustrate this, we present the same distribution as in Figure~\ref{img:parameters-cas}, but color-coded by the median $n_{\text{s\'ersic}}$ within each bin, as shown in Figure~\ref{img:parameters-cas-nsersic}.
In the Gini–M$_{20}$ diagrams, galaxies with smaller M$_{20}$ have higher $n_{\text{s\'ersic}}$.
Furthermore, a clear trend emerges that the disks tend to have median $n_{\text{s\'ersic}}< 2.5$, whereas early-type galaxies (E/S0/Sa) exhibit  $n_{\text{s\'ersic}>2.5}$.
Meanwhile, in the asymmetry–concentration panels, $n_{\text{s\'ersic}}$ increases monotonically with concentration.
These features underscore the strong correlation between non-parametric statistics and $n_{\text{s\'ersic}}$, consistent with the findings of \citet{Kartaltepe2023} based on a smaller sample.
We note that these distributions include all galaxies, encompassing a mixture of Hubble types. 
In Section~\ref{sec:gl_vs_ml}, we incorporate machine learning based morphological classifications to provide further insights.

In Figure~\ref{img:delta_gmc_z}, we further investigate the variation of non-parametric morphological statistics between the optical and UV, and how it evolves as a function of redshift. 
Interestingly, the median value $M_{20}$ is smaller in the rest-frame optical compared to the UV by about 0.15 at $z<6$.
The smaller $M_{20}$ values indicate that the spatial distribution of brightest galaxy light tends to be more concentrated in the optical compared to the UV.
Meanwhile, the concentration parameters are also slightly larger in the optical band.
On the other hand, galaxies appear more asymmetric in the optical than in the UV.
This may imply that dust is both abundant and patchily distributed at these redshifts, effectively smearing the UV light and reducing its apparent asymmetry.
For Gini and smoothness, there are generally no significant differences between the UV and optical measurements. 
At $z > 6$, the wavelength dependence of all those morphological parameters largely disappears, consistent with earlier JWST findings \citep[e.g.,][]{Treu2022}. 
This likely reflects the early stage of galaxy evolution in the high-redshift Universe, where galaxies have not yet had sufficient time for significant stellar aging, resulting in minimal differences between young and old stellar populations.

\section{Validation and comparison}\label{sec:comparsion}
In this Section, we compare our measurements to the \texttt{SE++} results and also those obtained from the machine learning technique. 
For the comparison, we consider secure sources with high SNR that are unaffected by the star mask (see Section~\ref{sec:results}).

\begin{figure*}
\centering
\includegraphics[width=1.8\columnwidth]{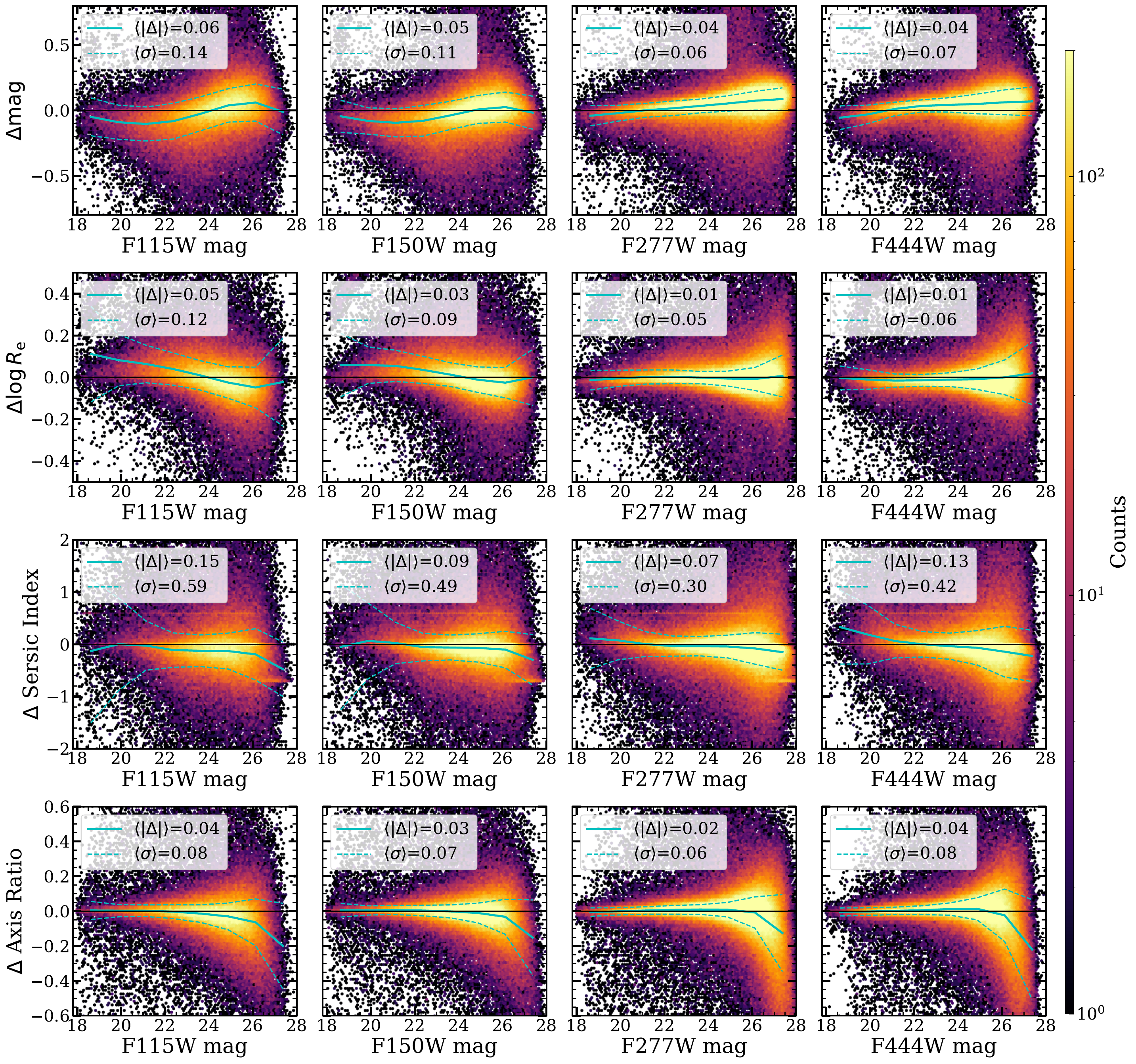}

\caption{S\'ersic parameters comparison between \texttt{Galight} and \texttt{SE++} as a function of \texttt{Galight} measured magnitude across four NIRCam bands with SNR $>10$ at each band. From the top to the bottom, the panels show the difference of magnitude, effective radius, S\'ersic index, and axis ratio obtained from single S\'ersic modeling between \texttt{Galight} and \texttt{SE++}.
In each panel, the solid cyan line indicates the median offset, while dashed line shows the scaled median absolute deviation (MAD) as a robust measure of scatter. }
\label{img:gl_vs_se}
\end{figure*}

\begin{figure*}
\centering
\includegraphics[width=1.8\columnwidth]{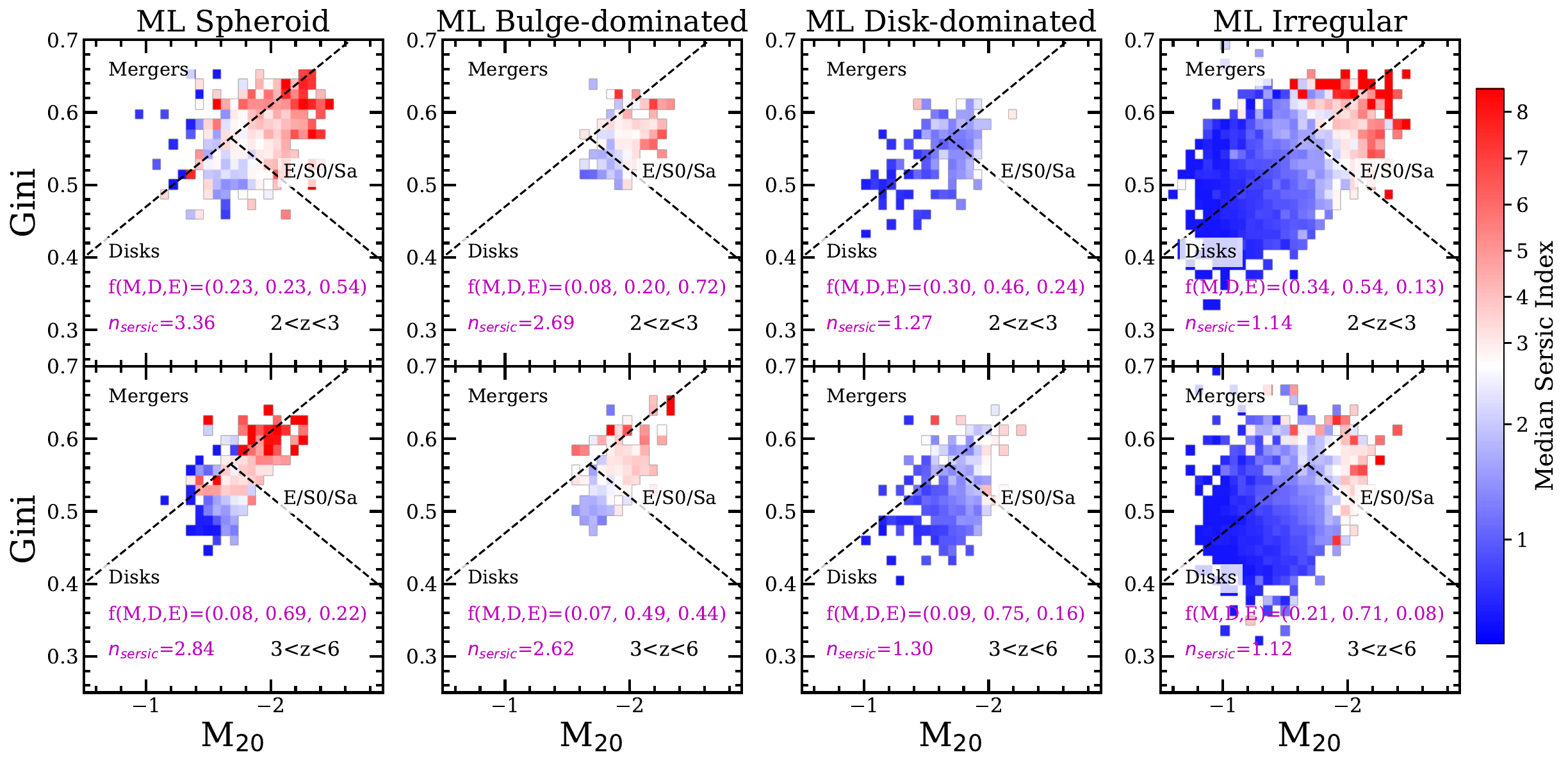}

\caption{The rest-frame optical Gini-M$_{20}$ distribution of four different morphological galaxies classified by machine learning technique \citep{Huertas-Company2025} with stellar mass $\log(M_*/M_\odot)>9.5$ at  $2<z<3$ (top) and $3<z<6$ (bottom).
The data points are color-coded by the median S\'ersic index within each bin, and the median S\'ersic index values for all sources in each morphological class are labeled in magenta. The fractions of mergers, disks, and elliptical galaxies $f(M, D, E)$ characterized by the Gini-M$_{20}$ diagram are labeled inside each panel, and the division lines are the same as in Figure~\ref{img:parameters-cas}.}

\label{img:cas-vs-ml}
\end{figure*}

\subsection{S\'ersic parameters comparison with SE++}\label{sec:gl_vs_se}

In Figure~\ref{img:gl_vs_se}, we compare the S\'ersic parameters measured by \texttt{Galight} and \texttt{SE++}. 
To better present the comparison to the relatively faint end, we select sources with SNR $>10$.
In each panel, the cyan solid and dashed lines denote the averaged absolute median value ($\Delta=\text{parameter}_{galight}-\text{parameter}_{SE++}$) and also the scatter of the offset $\langle \sigma \rangle $, respectively, with the values labeled at the top-left corner. 
Specifically, the median offset of magnitude is within $\sim0.1$ mag across all bands, with the averaged absolute value within $0.06$ mag and an averaged scatter less than $0.15$ mag.
At two shorter bands, F115W and F150W, the offset is the smallest at 24--25 mag, and increases towards both brighter and fainter end, i.e., the difference reaches $\sim0.1$ mag at $21$ mag.
The best agreement occurs in F277W and F444W bands, where the averaged absolute value is smaller than $0.05$ mag and the scatter also remains small, i.e., $\langle \sigma \rangle \leq 0.07$.

For the effective radius, the smallest offset is also found in the F277W and F444W bands, with a subtle median difference of 0.01 dex. 
Whereas the larger discrepancies appear in the shorter-wavelength bands. At F115W and F150W, sizes measured with \texttt{Galight} are larger than those from \texttt{SE++} by approximately 0.1 dex at magnitudes brighter than 24, but become smaller at fainter magnitudes (mag $> 24$). 
These discrepancies at shorter wavelengths may arise from uncertainties in the modeling approaches used by both methods. 
More importantly, the \texttt{SE++} morphological measurements represent average values over the 1--5 micron, which may contribute to the observed offset. Additionally, the longer-wavelength bands have deeper imaging depth,  contributing to the better agreement in F277W and F444W.

For both the S\'ersic index and axis ratio, there is a strong agreement between \texttt{Galight} and \texttt{SE++}. 
Median offsets in S\'ersic index across four bands are within $0.15$, with scatters smaller than 0.7. 
The axis ratios show excellent consistency, with noticeable differences emerging only at the faint end.

In summary, there is good overall agreement between \texttt{Galight} and \texttt{SE++} in terms of single S\'ersic modeling. 
However, some discrepancies remain in the measurements, which arise from differences in the methodology. 
\texttt{SE++} aims to deliver photometric measurements consistently across more than 30 bands, suppressing uncertainties due to morphological variation, whereas \texttt{Galight} performs independent measurements in each band, enabling the study of morphological changes as a function of wavelength.

\subsection{Comparison with Machine Learning results}\label{sec:gl_vs_ml}
Machine learning (ML) is another efficient method for galaxy morphological classification.
We compare our results with the machine learning catalog from \cite{Huertas-Company2025}, which applied a supervised convolutional neural network method \citep{Huertas-Company2024} to classify galaxies into four broad morphological classes: spheroids/pure bulges,  bulge-dominated, disk-dominated, and irregulars/disturbed/peculiar.
Additionally, the combination of the first and second classes, and the third and fourth classes are considered as early-type and late-type galaxies, respectively.
The machine learning catalog provides probability for each morphological class for COSMOS-Web galaxies in three NIRCam filters, F150W, F277W and F444W. 
Figure~\ref{img:cas-vs-ml} shows Gini-M$_{20}$ planes color-coded by the median S\'ersic index $n_{\text{s\'ersic}}$ for four morphological classes, respectively.
We use the F150W and F277W results for $2<z<3$ and $3<z<6$, respectively, corresponding to the rest-frame optical wavelength.
For each morphological class, we calculate the fraction of mergers, disks, and ellipticals diagnosed in Gini-M$_{20}$ plane.
Additionally, we also calculate the median value of S\'ersic index in each panel. 
The above two statistics are labeled at left-bottom inside each panel. 

At $2 < z < 3$ (top panels), the majority of ML early-type galaxies (spheroid and bulge-dominated) are classified as E/S0/Sa types, i.e., 54\% and 72\%, respectively. 
The median $n_{\text{s\'ersic}}$ are 3.36 and 2.69. 
In contrast, ML late-type galaxies (disk-dominated and irregular) are primarily populated by disks, with approximately half of the sources comprising disks. The median  $n_{\text{s\'ersic}}$ values for these populations are 1.27 and 1.14, consistent with typical disk-like light profiles.
At $3 < z < 6$ (bottom panels), the median $n_{\text{s\'ersic}}$ are 2.84, 2.62, 1.30, and 1.12 for ML spheroids, bulge-dominated, disk-dominated, and irregular classes, respectively, which are close to the values at lower redshift ($2<z<3$). 
However, despite this structural consistency in S\'ersic index, the structure features classified via non-parametric statistics change, with the fraction of `Disks' dominating all four morphological classes. For example, ML spheroids consist of 69\% disks and only 22\% E/S0/Sa. Although \citet{Huertas-Company2024} clarified that the spheroid classification typically refers to round and compact galaxies, which do not necessarily correspond to kinematically hot systems, particularly at high redshift. 
For ML late-type galaxies, the fraction of disks reaches 75\%, with only a small fraction classified as ellipticals.

According to the comparison results, there is also a good agreement between the results obtained in this work and the morphological classification adopted from machine learning, that all ML early-type galaxies 
have $n_{\text{s\'ersic}}>2.5$, whereas ML late-type galaxies have $n_{\text{s\'ersic}}\sim1.2$.
For Gini-M$_{20}$ statistics, a generally good agreement is also observed at $2<z<3$, with early-type galaxies dominated by ellipticals and late-type galaxies dominated by disks. 
At higher redshifts, however, disks dominate all morphological types, highlighting unique structural properties that non-parametric statistics can reveal.

\begin{figure*}
\includegraphics[width=0.65\columnwidth]{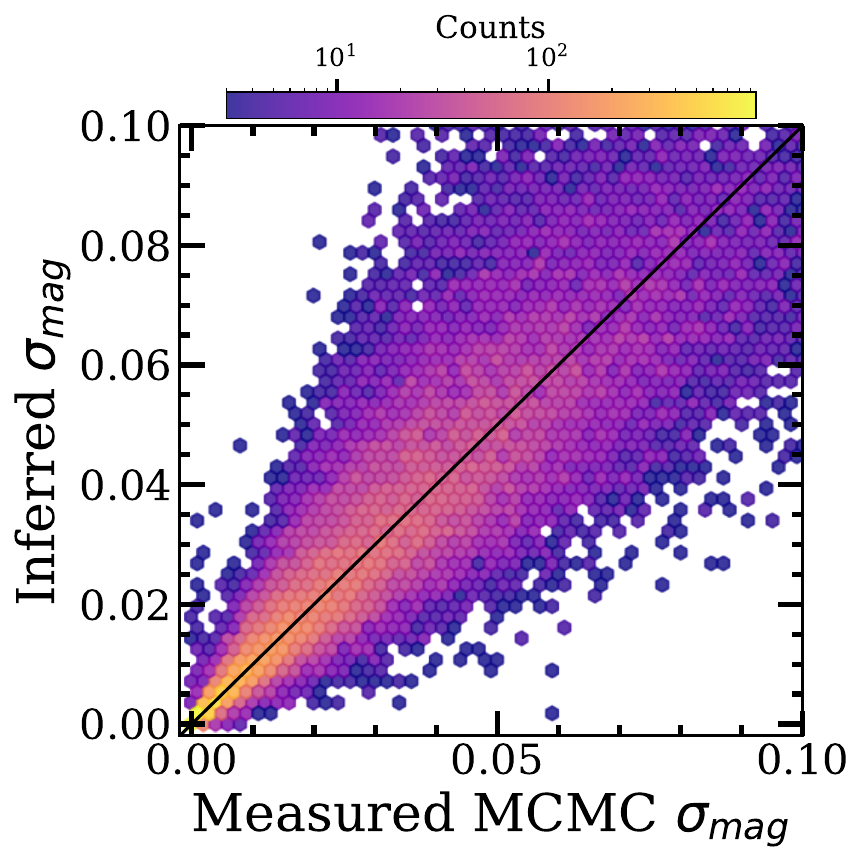}
\includegraphics[width=0.65\columnwidth]{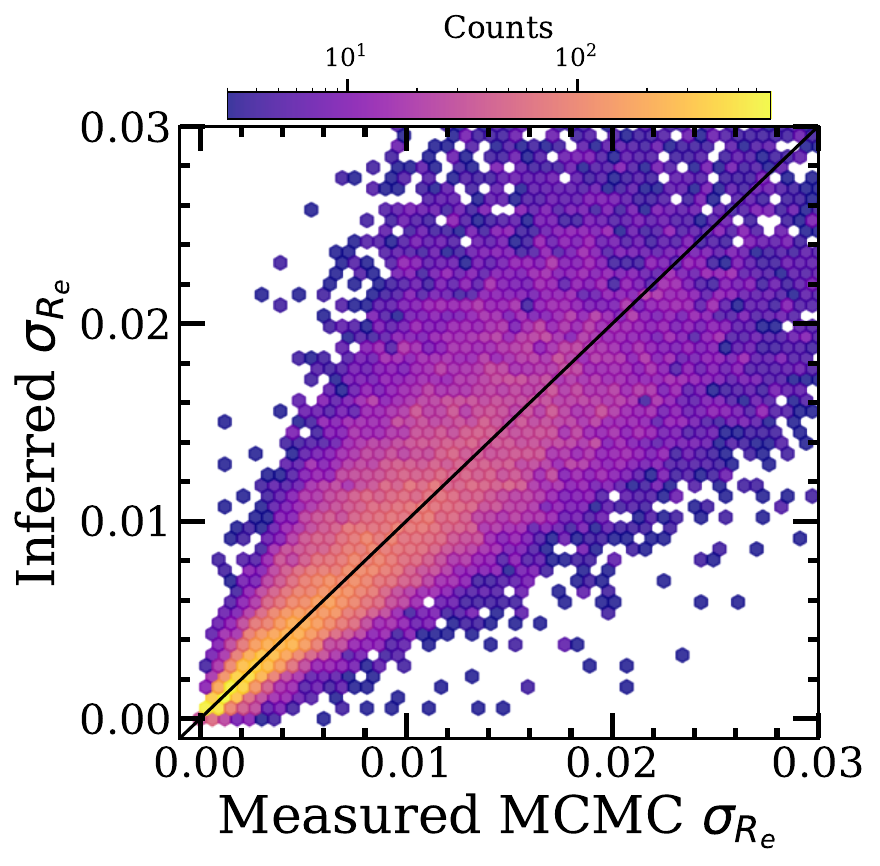}
\includegraphics[width=0.65\columnwidth]{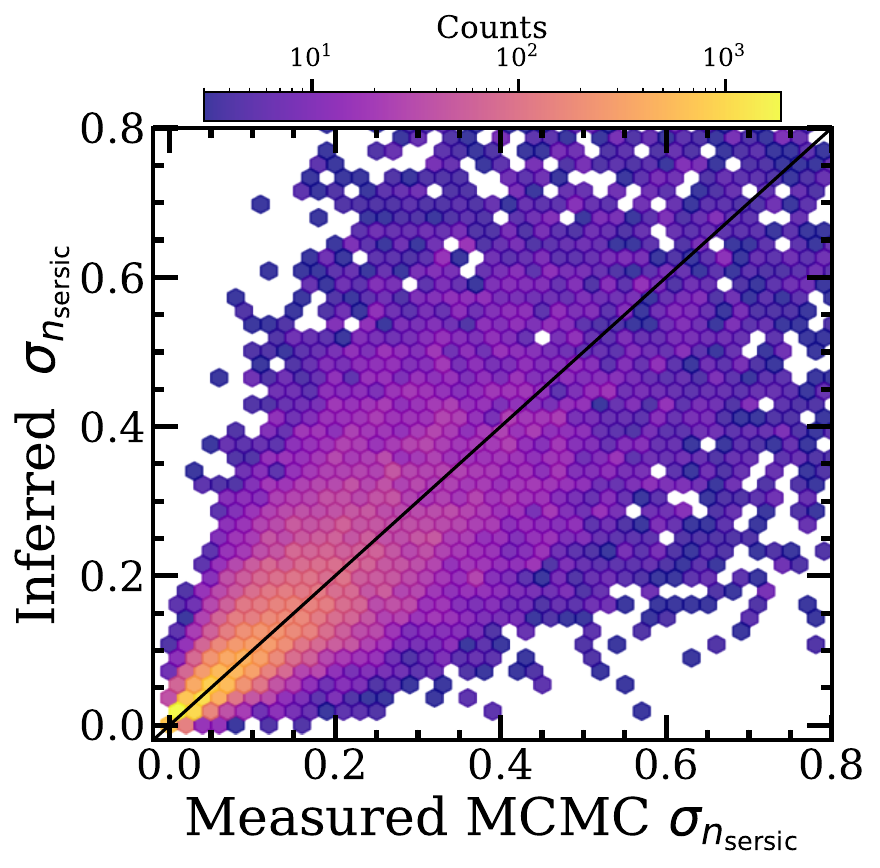}
\caption{Comparison between MCMC-derived measurement uncertainties and those inferred from the empirical nearest neighbor method, for magnitude (left), half-light radius $R_e$ (middle), and S\'ersic index $n_\text{s\'ersic}$ (right).}
\label{img:comparison_mcmc_err}
\end{figure*}

\section{uncertainties}\label{sec:uncertainties}
\subsection{Measurement uncertainty}
We have adopted MCMC methods to estimate the measurement uncertainties of parametric fittings.
However, the full MCMC fitting for all sources in the COSMOS2025 catalog is computationally expensive.
Therefore, we adopt an empirical approach to handle the large amount of data. 
In brief, the key idea is to assign uncertainties to each source by averaging those of its nearest neighbors that have MCMC-derived uncertainties. The selection of nearest neighbors is based on similarity in key observables. The details are described below.

We first construct a reference sample by randomly selecting 50,000 sample galaxies and performing MCMC in each NIRCam filter to obtain posterior-based uncertainties for each parameter. 
Specifically, we take the initial estimation from PSO minimization as a starting point for the MCMC sampling process.
The $1\sigma$ distribution of the posterior distribution is attributed as their statistical uncertainty. 
We then identify the most similar galaxies in the reference sample for each galaxy. 
As inspired by \cite{vdW2012}, we compute the Euclidean distances between each target galaxy $p_t$ and all reference samples $p_i$ in the three-dimensional parameter space spanned by magnitude, half-light radius and S\'ersic index ($mag$, $R_e$, and $n_\text{s\'ersic}$).
For example, we define the $p_i$ as,
\begin{equation}
\begin{split}
 p_i = (mag_i/\sigma(mag), \log R_{e,i}/\sigma(\log R_e), \\
 \log n_{\text{s\'ersic},i}/ \sigma(\log n_\text{s\'ersic}));
 i=1, 2, 3 \cdots 50000,
\end{split}
\end{equation}
where $\sigma$ denotes the standard deviation in the reference parameters.
For each target galaxy $p_t$, we identify its 25 nearest neighbors in the reference sample.
Because the uncertainty also depends on SNR, we account for this by scaling the MCMC-based uncertainties of the reference galaxies by their SNR values. 
The final uncertainty for each target galaxy is then estimated as the median of the scaled MCMC uncertainties of its neighbors, divided by the SNR of the target.
We verified that varying the number of selected neighbors (e.g, 100 or 200) does not significantly affect the results. 

As a result, we compare the inferred MCMC uncertainties from the above method to those obtained from actual MCMC runs, as shown in Figure~\ref{img:comparison_mcmc_err}.
There is a great consistency between measured and inferred uncertainties for S\'ersic parameters, which validates our method for assigning uncertainties.

\subsection{Systematic simulation}
We further conduct galaxy injection simulations to assess the accuracy and precision of our structural measurement. 
We mock galaxy light with a single S\'ersic model and then convolve with the same PSF used in data analysis. The S\'ersic parameters are randomly generated based on the distribution of parameter values observed in the real data. These simulated galaxies are then injected into blank regions of actual COSMOS-Web imaging mosaics.
We measured the structural parameters of the simulated galaxies using the same techniques applied to the real data. 
Additionally, to account for uncertainties arising from the PSF, we fit the simulated sources using a different PSF. 
By comparing the input and output values, we assess the offset and scatter for  S\'ersic parameters.

\begin{figure}
\centering
\includegraphics[width=0.9\columnwidth]{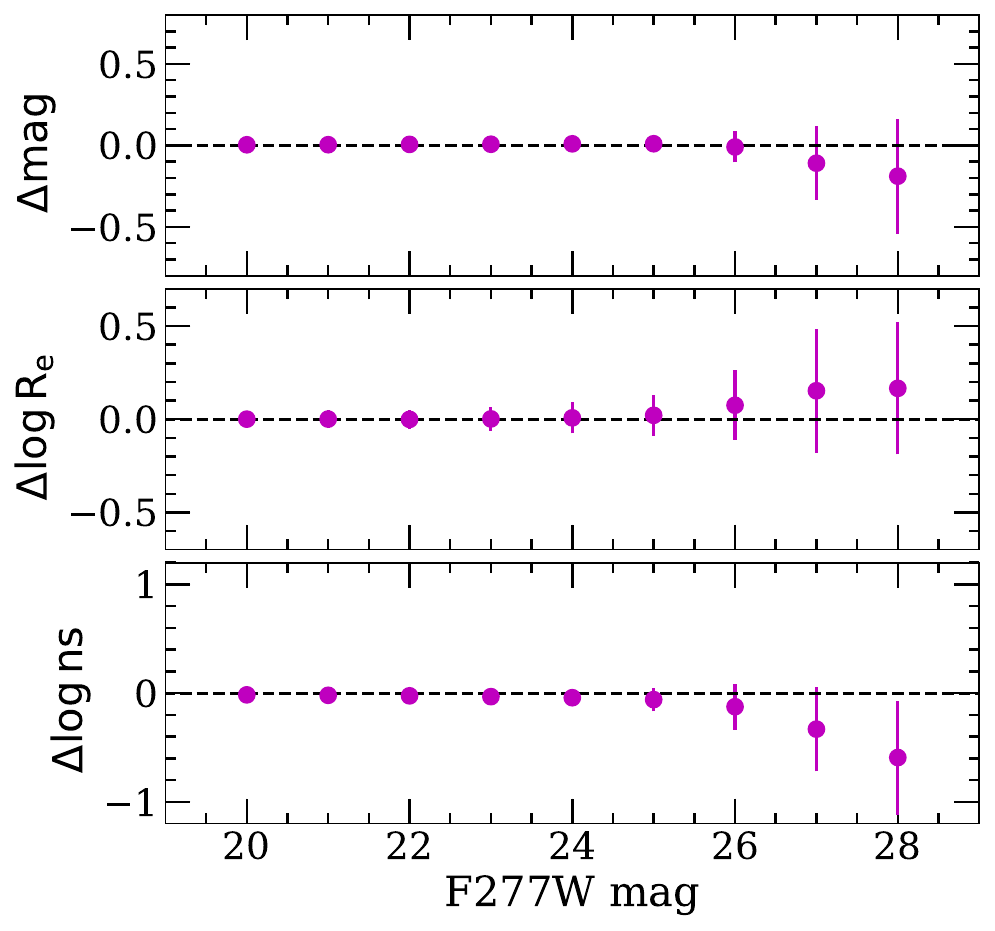}
\caption{The difference of magnitude, half-light radius, and S\'ersic index between the output and input derived from the simulation dataset as a function of output magnitude.
}
\label{img:simulation_dr_mag_ns_F277W}
\end{figure}

Figure~\ref{img:simulation_dr_mag_ns_F277W} presents the median offsets and their 1$\sigma$ scatter of magnitude, half-light radius, and S\'ersic index between the input and output half-light radius, as a function of magnitude.
The top panel shows the $\Delta mag$ that remains close to zero across the entire magnitude range with some offset at fainter magnitudes, e.g., $\Delta mag\sim$0.1-0.2 at magnitudes fainter than 27 mag. 
The middle panel shows that the $\Delta\log R_e$ increases slightly towards faint-end, indicating overestimation of size for fainter galaxies. 
We find the median offset is within 0.1 dex for sources brighter than 26.5 mag, and within 0.2 dex even towards 28 mag.
This result confirms that our size measurements are robust for galaxies, especially for those with magnitudes brighter than 26.5.
Similarly, the S\'ersic index differences as shown in bottom panel remain relatively stable but exist offset at fainter magnitudes. 

\begin{figure}
\centering
\includegraphics[width=0.9\columnwidth]{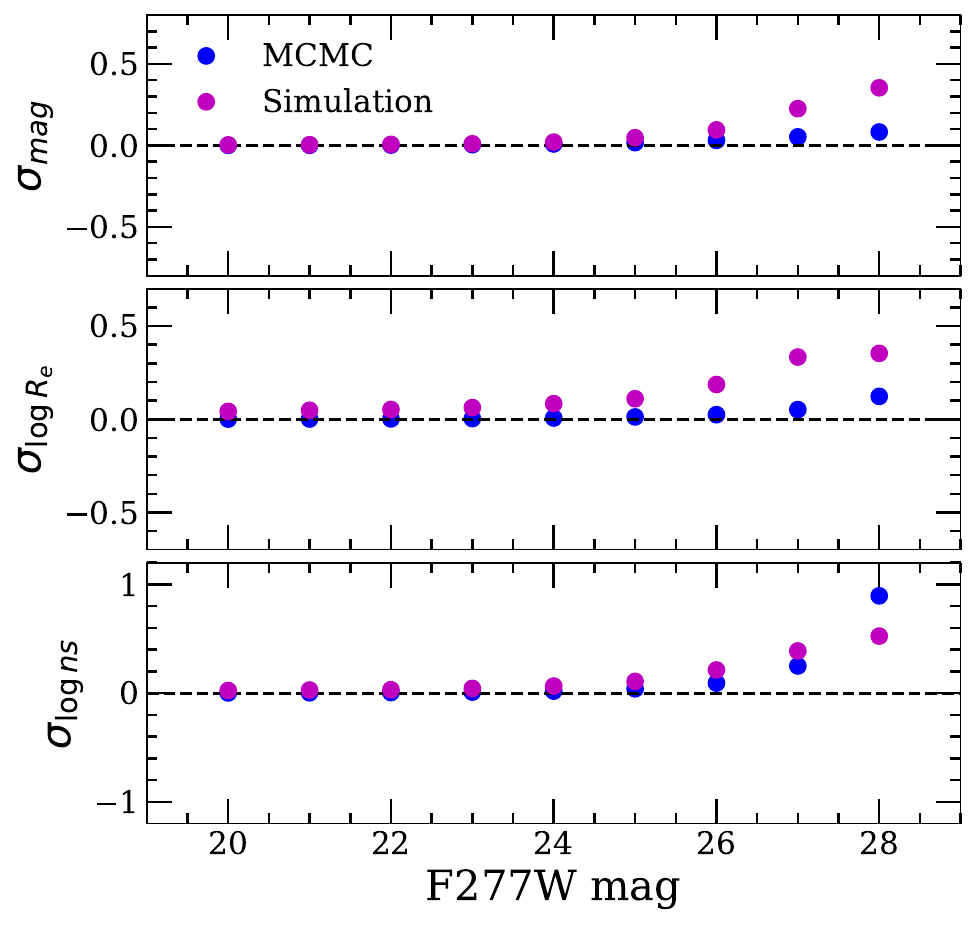}

\caption{Comparison of uncertainties in magnitude, half-light radius, and S\'ersic index derived from simulation and MCMC-based analysis. 
Magenta points represent the 1$\sigma$ scatter of the difference between the output and input sizes in the simulation. 
Blue points represent the uncertainties derived from the MCMC posterior distribution in real source measurement.
}
\label{img:uncertainty_logre_sim_vs_mcmcF277W}
\end{figure}

Besides the offset, the scatter in $\Delta mag$, $\Delta\log R_e$ and 
$\Delta \log n_{\text{s\'ersic}}$ reflects the accuracy of reproducing a galaxy modeled with a S\'ersic profile, and we compare it with the uncertainty derived from MCMC.
This comparison is shown in Figure~\ref{img:uncertainty_logre_sim_vs_mcmcF277W}.
As expected, the uncertainties obtained from simulation (magenta points) is generally larger than that from MCMC (blue points), because the scatter in simulation captures a broader range of measurement uncertainties, including unmodeled random background noise, PSF mismatch, and statistical uncertainties reflected in the MCMC posterior.
In contrast, the MCMC uncertainty reflects only statistical uncertainties under the assumed model.
We find that the MCMC uncertainty  
is comparable to that obtained from simulation for galaxies brighter than 25 mag, but becomes much smaller at the faint end for both magnitude and half-light radius.
For example, the MCMC uncertainties $\sigma_{\log R_e}$ remain $<0.1$ dex across entire magnitude range, whereas the uncertainties from the simulation reach  
0.1 dex at 26 mag and exceed 0.3 dex at fainter magnitudes.    
In summary, for most galaxies in our sample, MCMC provides reasonably reliable uncertainty estimates.
However, it may underestimate uncertainty for faint sources.
In our catalog, we report only the MCMC-derived uncertainties for each parameter, and recommend that users account for potential systematic offsets and additional uncertainty depending on their specific scientific applications.

\section{Summary \& Conclusion}\label{sec:sum_and_conclusion}
In this work, we present multi-wavelength morphological measurements for $\sim780,000$ galaxies in the COSMOS2025 catalog.
We perform both parametric and non-parametric analyses independently on four NIRCam filters. 
For the parametric fits, we adopt three modeling strategies, (1) a single S\'ersic profile, (2) a double S\'ersic profile with separated disk and bulge components, and (3) a point-source decomposition. 
For the non-parametric analyses, we measure a set of widely used indicators, including the Gini coefficient, $M_{20}$, concentration, asymmetry, and smoothness/clumpiness.
Table~\ref{table:cata} describes the columns that included in our morphology catalog.

Leveraging the COSMOS2025 catalog that includes photometric redshifts and physical parameters such as stellar mass and star-formation rate, we examine the evolution of structural parameters and their dependence on stellar mass and rest-frame wavelength over $2<z<10$.
For the parametric fits, we find that:
\begin{itemize}
\item{\em
A strong correlation between galaxy structure and star formation activity is evident from the distribution of $n_{\text{s\'ersic}}$ in the SFR--M$_*$ planes.}
Galaxies with
exponential profiles ($n_{\text{s\'ersic}} \sim 1$) are closely following the main sequence of star-forming galaxies up to $z \sim 4$.
At higher redshifts, especially at $z > 4$, the correlation between $n_{\text{s\'ersic}}$ and position on the main sequence becomes less clear. 
Interestingly, a fraction of high-redshift quiescent galaxies still exhibit disk-like structures ($n_{\text{s\'ersic}} \leq 1$).
Furthermore, we find the evolution of $n_{\text{s\'ersic}}$ across redshift bins ($2 < z < 10$) shows clear dependence on stellar mass. For the most massive galaxies ($M_* > 10^{10.5} M_{\odot}$), the median $n_{\text{s\'ersic}}$ increases from $\sim 1$ at $z \sim 6$ to $\sim 2.5$ at $z \sim 2$, indicating a trend toward more compact, bulge-dominated structures with cosmic time.
At lower stellar masses, the evolution is less pronounced, with $n_{\text{s\'ersic}}$ remaining relatively constant around $1.2$ at all redshifts
Additionally, a decrease in the median $n_{\text{s\'ersic}}$ at $z > 4$ for lower mass galaxies suggests an increasing prevalence of irregular galaxies at higher redshifts, which are more common in the lower stellar mass range.
The UV $n_{\text{s\'ersic}}$ is systematically smaller than its in optical, but their evolution show similar trends. 

\item {\em The distribution of $B/T$ on SFR--M$_*$ plane and evolution of $B/T$ follow similar trends to those of S\'ersic index.}
The evolution of the $B/T$ ratio is also mass-dependent.
There is little evolution for galaxies with $\log(M_*/M_{\odot})<10.5$ at $2<z<10$, but it increases from $\sim20\%$ at $z\sim5$ to $\sim30\%$ at $z\sim2$ for the most massive galaxies bin with $10.5<\log(M_*/M_{\odot})<11$.
There is a tight correlation between the S\'ersic index and bulge-to-total ratio ($B/T$), with $n_{\text{s\'ersic}}\sim1$ corresponding to median $B/T\sim10\%$ and $n_{\text{s\'ersic}}\sim2.5$ reaching $B/T\sim50\%$, consistent across all NIRCam bands. 

\item {\em AGN hosts have higher S\'ersic index values.}
From the distribution of the S\'ersic index obtained from the decomposition of a Single S\'ersic and a central point source,
S\'ersic index of X-ray detected sources exhibit systematically higher value than that of all galaxies, with a pronounced excess at $n_{\text{s\'ersic}}>2$ across all NIRCam bands.

\end{itemize}

For non-parametric fits, we have presented distributions of galaxies on Gini–M$_{20}$ and asymmetry–concentration planes, and examined their correlation with S\'ersic index obtained from parametric fitting. 
\begin{itemize}

\item We find these non-parametric diagnostics show strong correlations with the S\'ersic index, such as disk galaxies typically have $n_{\text{s\'ersic}}<2.5$, while early-type galaxies show higher values.
We have also investigated the variation of non-parametric indicators as a function of wavelength, and found that in the rest-frame optical, galaxies are more concentrated and less asymmetric compared to the UV. 
The variations are observed at $z\sim6$, and the wavelength dependence largely disappears at higher redshift.

\end{itemize}

We have validated our measurements against \texttt{SE++}, finding overall good agreement in S\'ersic parameters across all NIRCam bands, with magnitude offsets within 0.1 mag, half-light radius offsets $\lesssim0.1$ dex, and consistent S\'ersic index and axis ratio estimates. 
Discrepancies at shorter wavelengths likely reflect differences in modeling approaches and averaging schemes. 
Moreover, comparison with ML classifications further confirms the reliability of our results that galaxies with $n_{\text{s\'ersic}}>2.5$ predominantly correspond to ML early-type classes, while those with $n_{\text{s\'ersic}}\sim1.1$ are associated with ML late-types.

Beyond the results presented in this work, our morphological catalog offers a rich and comprehensive dataset of multi-wavelength structural measurements that has also been used in many recent studies \citep{Yang2025, Ding2025, Gozaliasl2025}.
It provides a foundation for a wide range of future studies, including bulge growth, the impact of AGN activity, and structural transformations across cosmic time. 
We expect that this catalog will serve as a valuable resource for the community, facilitating new insights and constraints in galaxy formation and evolution.

\startlongtable
\begin{deluxetable*}{cc}
\tablecaption{Morphology Catalog Column Descriptions\label{table:cata}}
\tablehead{
\colhead{Column Name} & \colhead{Description} 
}
\startdata
\hline
ID & ID same as the COSMOS-Web photometry catalog \\
RA & Right ascension\\
DEC & Declination\\
rearc\_xxx\_sersic & Half-light radius (arcsecond) of S\'ersic model in xxx filter\\
nsersic\_xxx\_sersic & S\'ersic index of S\'ersic model in xxx filter\\
phi\_xxx\_sersic & Position angle of S\'ersic model in xxx filter\\
qratio\_xxx\_sersic & Axis ratio of S\'ersic model in xxx filter \\
mag\_xxx\_sersic & Magnitude of S\'ersic model in xxx filter\\
rearc\_xxx\_sersic\_err & Half-light radius err (arcsecond) of S\'ersic model in xxx filter\\
nsersic\_xxx\_sersic\_err & S\'ersic index err of S\'ersic model in xxx filter\\
phi\_xxx\_sersic\_err & Position angle err of S\'ersic model in xxx filter\\
qratio\_xxx\_sersic\_err & Axis ratio err of S\'ersic model in xxx filter\\
mag\_xxx\_sersic\_err & Magnitude err of S\'ersic model in xxx filter\\
rearc\_bulge\_xxx\_bd & Half-light radius (arcsecond) of bulge component in xxx filter\\
nsersic\_bulge\_xxx\_bd & S\'ersic index of bulge component in xxx filter\\
phi\_bulge\_xxx\_bd & Position angle of  bulge component in xxx filter\\
qratio\_bulge\_xxx\_bd & Axis ratio of  bulge component in xxx filter\\
mag\_bulge\_xxx\_bd & Magnitude of  bulge component in xxx filter\\
rearc\_disk\_xxx\_bd & Half-light radius (arcsecond) of disk component in xxx filter\\
nsersic\_disk\_xxx\_bd & S\'ersic index of  disk component in xxx filter\\
phi\_disk\_xxx\_bd & Position angle of disk component in xxx filter\\
qratio\_disk\_xxx\_bd & Axis ratio of  disk component in xxx filter\\
mag\_disk\_xxx\_bd & Magnitude of disk component in xxx filter\\
rearc\_bulge\_xxx\_bd\_err & Half-light radius err (arcsecond) of bulge component in xxx filter\\
phi\_bulge\_xxx\_bd\_err & Position angle of err bulge component in xxx filter\\
qratio\_bulge\_xxx\_bd\_err & Axis ratio err of  bulge component in xxx filter\\
mag\_bulge\_xxx\_bd\_err & Magnitude err of  bulge component in xxx filter\\
rearc\_disk\_xxx\_bd\_err & Half-light radius err (arcsecond) of disk component in xxx filter\\
phi\_disk\_xxx\_bd\_err & Position angle err of disk component in xxx filter\\
qratio\_disk\_xxx\_bd\_err & Axis ratio err of  disk component in xxx filter\\
mag\_disk\_xxx\_bd\_err & Magnitude err of disk component in xxx filter\\
rearc\_host\_xxx\_ps &  Half-light radius of extended component in xxx filter\\
nsersic\_host\_xxx\_ps & S\'ersic index of extended component in xxx filter\\
phi\_host\_xxx\_ps & Position angle of extended component extended in xxx filter\\
qratio\_host\_xxx\_ps & Axis ratio of  extended component in xxx filter\\
mag\_host\_xxx\_ps & Magnitude of extended component in xxx filter\\
p2t\_flux\_ratio\_xxx\_ps & Point to total flux ratio in xxx filter\\
rearc\_host\_xxx\_ps\_err &  Half-light radius err (arcsecond) of extended component in xxx filter\\
nsersic\_host\_xxx\_ps\_err & S\'ersic index err of extended component in xxx filter\\
phi\_host\_xxx\_ps\_err &  Position angle err of extended component extended in xxx filter\\
qratio\_host\_xxx\_ps\_err& Axis ratio err of extended component in xxx filter\\
mag\_host\_xxx\_ps\_err & Magnitude err of extended component in xxx filter\\
asymmetry\_xxx & Asymmetry in xxx filter\\
smoothness\_xxx & Smoothness in xxx filter\\
concentration\_xxx & Concentration in xxx filter\\
gini\_xxx & Gini in xxx filter\\
m20\_xxx & M$_{20}$ in xxx filter\\
cas\_flag\_xxx & \texttt{Statmorph} quality flag\\
bic\_xxx\_sersic & Bayesian information criterion of S\'ersic model\\
reduced\_Chisq\_xxx\_sersic & Reduced $\chi^2$ of S\'ersic model\\
bic\_list\_xxx\_bd &  Bayesian information criterion of Bulge+Disk model\\
reduced\_Chisq\_list\_xxx\_bd & Reduced $\chi^2$ of Bulge+Disk model\\
bic\_list\_xxx\_ps &  Bayesian information criterion of Point+Extended model\\
reduced\_Chisq\_list\_xxx\_ps & Reduced $\chi^2$ of Point+Extended model\\
\enddata
\end{deluxetable*}

\section{acknowledgments}
This work is based on observations made with the NASA/ESA/CSA James Webb Space Telescope. The data were obtained from the Mikulski Archive for Space Telescopes at the Space Telescope Science Institute, which is operated by the Association of Universities for Research in Astronomy, Inc., under NASA contract NAS 5-03127 for JWST. These observations are associated with program \#1727. Support for this work was provided by NASA through grant JWST-GO-01727 awarded by the Space Telescope Science Institute, which is operated by the Association of Universities for Research in Astronomy, Inc., under NASA contract NAS 5-26555.
This project has received funding from the European Union’s Horizon
2020 research and innovation programme under the Marie
Skłodowska-Curie grant agreement No 101148925.
This work was made possible by utilizing the CANDIDE cluster at the Institut d'Astrophysique de Paris. The cluster was funded through grants from the PNCG, CNES, DIM-ACAV, the Euclid Consortium, and the Danish National Research Foundation Cosmic Dawn Center (DNRF140). It is maintained by Stephane Rouberol. Some of the measurements in this work are supported by World Premier International Research Center Initiative (WPI Initiative), MEXT, Japan.

\bibliography{sample631}{}
\bibliographystyle{aasjournal}

\end{document}